\PassOptionsToPackage{utf8}{inputenc}
\documentclass[nocrop,letter]{bioinfo}
\DeclareUnicodeCharacter{2212}{-}
\copyrightyear{2019} \pubyear{2019}
\pdfoutput=1

\access{Advance Access Publication Date: Day Month Year}
\appnotes{Manuscript Category}
\usepackage{amsmath,bm}
\usepackage{multirow}
\usepackage{siunitx}
\usepackage{graphicx}
\usepackage{color}
\usepackage{subcaption}
\usepackage[justification=centering]{caption}
\usepackage{longtable}

\newcommand{\norm}[1]{\left\lVert#1\right\rVert}

\begin{document}

\setlength{\belowdisplayskip}{0pt} \setlength{\belowdisplayshortskip}{0pt}
\setlength{\abovedisplayskip}{0pt} \setlength{\abovedisplayshortskip}{0pt}

\firstpage{1}

\subtitle{Structural Bioinformatics}

\title[Bayesian active learning for protein docking]{Bayesian active learning for optimization and uncertainty quantification in protein docking}

\author[Cao and Shen]{Yue Cao\ and Yang Shen\,$^{\ast}$}
\address{Department of Electrical and Computer Engineering, Texas A\&M University, College Station, TX 77843, United States.}

\corresp{$^\ast$To whom correspondence should be addressed.}

\history{Received on XXXXX; revised on XXXXX; accepted on XXXXX}

\editor{Associate Editor: XXXXXXX}

\vspace{-2em}
\abstract{\textbf{Motivation:} \textit{Ab initio} protein docking represents a major challenge for optimizing a noisy and costly ``black box"-like function in a high-dimensional space. Despite progress in this field, there is no docking method available for rigorous uncertainty quantification (UQ) of its solution quality (e.g. interface RMSD or iRMSD). 
\\
\textbf{Results:} We introduce a novel algorithm, Bayesian Active Learning (BAL), for optimization and UQ of such black-box functions and flexible protein docking. BAL directly models the posterior distribution of the global optimum (or native structures for protein docking) with active sampling and posterior estimation iteratively feeding each other. Furthermore, we use complex normal modes to represent a homogeneous Euclidean conformation space suitable for high-dimension optimization and construct funnel-like energy models for encounter complexes. Over a protein docking benchmark set and a CAPRI set including homology docking, we establish that BAL significantly improve against both starting points by rigid docking and refinements by particle swarm optimization, providing for one third targets a top-3 near-native prediction. BAL also generates tight confidence intervals with half range around 25\% of iRMSD and confidence level at 85\%.  Its estimated probability of a prediction being native or not achieves binary classification AUROC at 0.93 and AUPRC over 0.60 (compared to 0.14 by chance); and also found to help ranking predictions. To the best of our knowledge, this study represents the first uncertainty quantification solution for protein docking, with theoretical rigor and comprehensive assessment.  \\
\\
\textbf{Availability:} Source codes are available at {https://github.com/Shen-Lab/BAL}.\\
\textbf{Contact:} \href{yshen@tamu.edu}{yshen@tamu.edu}\\
\textbf{Supplementary information:} {https://github.com/Shen-Lab/BAL/tree/master/Paper\_SI/} }

\maketitle

\section{Introduction}

Protein-protein interactions underlie many cellular processes, which has been increasingly revealed by the quickly advancing high-throughput experimental methods.  However, compared to the binary information about what protein pairs interact, the structural knowledge about how proteins interact remains relatively scarce \citep{mosca2013interactome3d}. Protein docking helps close such a gap by computationally predicting the 3D structures of protein-protein complexes given individual proteins' 1D sequences or 3D structures \citep{smith2002prediction}. 

\textit{Ab initio} protein docking is often recast as an energy (or other objective functions) optimization problem.  For the type of objective functions relevant to protein docking, neither the analytical form nor the gradient information would help global optimization as the functions are non-convex and extremely rugged thus their gradients are too local to inform global landscape. So the objective functions are often treated as \textit{de facto} ``black-box'' functions for global optimization. Meanwhile, these functions are very expensive to evaluate. Various methods, especially refinement-stage methods, have progressed to effectively sample the high-dimensional conformational space against the expensive functional evaluations \citep{gray2003protein,moal_swarmdock_2010,shen2013improved,jimenez2017lightdock,marze_efficient_2018,pfeiffenberger2018refinement}.

When solving such optimization problems still remains a great challenge, quantifying the uncertainty of protein-docking results in some quality of interest (very often interface RMSD or iRMSD) is not addressed by any protein-docking method. Even though such information is much needed by the end users, current protein-docking methods often generate a rank-ordered list of results without giving quality estimation and uncertainty quantification to individual results and without providing the confidence in whether the entire list contains a quality result (for instance, a near-native protein-complex model with iRMSD $\leqslant$ 4\AA).  

Sources of uncertainty in protein docking methods include the objective function as well as the sampling scheme, which can be classified as epistemic uncertainty \citep{UQSources}. For instance, energy models as objective functions provide noisy and approximate observations of the assumed ground truth --- the Gibbs free energy; and iterative sampling techniques suffer from both the approximation of the search space (e.g. rotamerized side chains) and insufficient data in the approximated space (e.g. small numbers of samples considering the high dimensionality of the search space).  In addition, uncertainty in protein structure data (e.g. X-ray crystal structures of proteins being "averaged" versions of their native conformations and derived from fitting observed diffraction patterns), which can be classified as aleatoric uncertainty, also enters protein docking methods when crystal structures are used as ground-truth native structures for training objective functions or tuning parameters in protein-docking methods. 

Whereas the forward propagation of aleatoric uncertainty in protein structure data to structure-determined  quantities has been studied empirically \citep{li_estimation_2016} and theoretically \citep{Rasheed17}, the much more difficult, inverse quantification of uncertainty in predicted protein or protein-complex structures originating from epistemic uncertainty in computational methods, is still lacking a mathematically rigorous solution.  A unique challenge for uncertainty quantification (UQ) in protein docking is that the desired quality of interest here is directly determined by the optimum itself rather than the optimal value. In other words, closeness to native structures (for instance, measured by iRMSD) is an indicator for the usefulness of the docking results, but closeness to native structures' energy values is not necessarily the case.  Therefore, UQ in protein docking has to be jointly solved with function optimization when finding the inverse mapping from a docking objective function to its global optimum is neither analytically plausible nor empirically cheap.

In this study, we introduce a rigorous Bayesian framework to simultaneously perform function optimization and uncertainty quantification for expensive-to-evaluate black-box objective functions.  To that end, our Bayesian active learning (BAL) iteratively and adaptively generate samples and update posterior distributions of the global optimum. We propose a posterior in the form of the Boltzmann distribution with adaptive annealing schedule that gradually shifts the search from exploration to exploitation and with an objective-function estimator (surrogate) that is a non-parametric Kriging regressor.  The iteratively updated posterior carries the belief (and uncertainty as well) on where the global optimum is given historic samples and guides next-iteration samples, which presents an efficient data-collection scheme for both optimization and UQ. Compared to typical Bayesian optimization methods \citep{shahriari_taking_2016} that first model the posterior of the objective function and then optimize the resulting functional, our BAL framework directly models the posterior for the global optimum and overcomes the intensive computations in both steps of typical Bayesian optimization methods.  Compared to another work \citep{ortega_nonparametric_2012} that also models the posterior of the global optimum, we provide both theoretical and empirical results that our BAL has a consistent and unbiased estimator as well as a global uncertainty-aware and dimension-dependent annealing schedule.    

We also make innovative contributions in the application domain of protein docking.  Specifically, we design a machine learning-based objective function that estimates binding affinities for docked encounter complexes as well as assesses the quality of interest, iRMSD, for docking results. We also re-parameterize the search space for both external rigid-body motions \citep{shen2008protein} and internal flexibility, into a low-dimensional homogeneous and isotropic space suitable for high-dimensional optimization, using our (protein) complex normal modes (cNMA) \citep{oliwa_cnma:_2015}. Considering that protein docking refinement often starts with initial predictions representing separate conformational clusters/regions, we use estimated local posteriors over individual regions to construct local and global partition functions; and then calculate the probability that the  prediction for each conformational cluster, each conformational cluster, or the entire list of conformational clusters is near-native.

The rest of the paper is organized as following.  In Materials and Methods, we first give a mathematical formulation for the optimization and the UQ, then introduce our Bayesian active learning (BAL) that iteratively update sampling and posterior estimation. We next introduce the parameterization of the search space that allows concurrent and homogeneous sampling of external rigid-body and internal flexible-body motions as well as newly-developed machine learning models as the noisy energy function that estimates the binding free energy for encounter complexes. And we end the Materials and Methods with uncertainty quantification for protein docking.  In Results and Discussion, using a comprehensive protein docking benchmark set involving unbound docking and a CAPRI set involving homology docking, we assess optimization results for BAL with comparison to starting structures from ZDOCK and refined structures by particle swarm optimization (PSO). We further assess the uncertainty quantification results:  accuracy of the confidence levels  and tightness of the confidence region, as well as the confidence scores on the near-nativeness of predictions. Lastly, before reaching Conclusions, we visualize the estimated energy landscape and confirm that the funnel-like energy landscapes do exist near native structures in the homogeneous conformational space blending external rigid-body and internal flexible-body motions.

\begin{figure*}
   \centering
    \begin{subfigure}[b]{0.23\textwidth}
        \includegraphics[width=\textwidth]{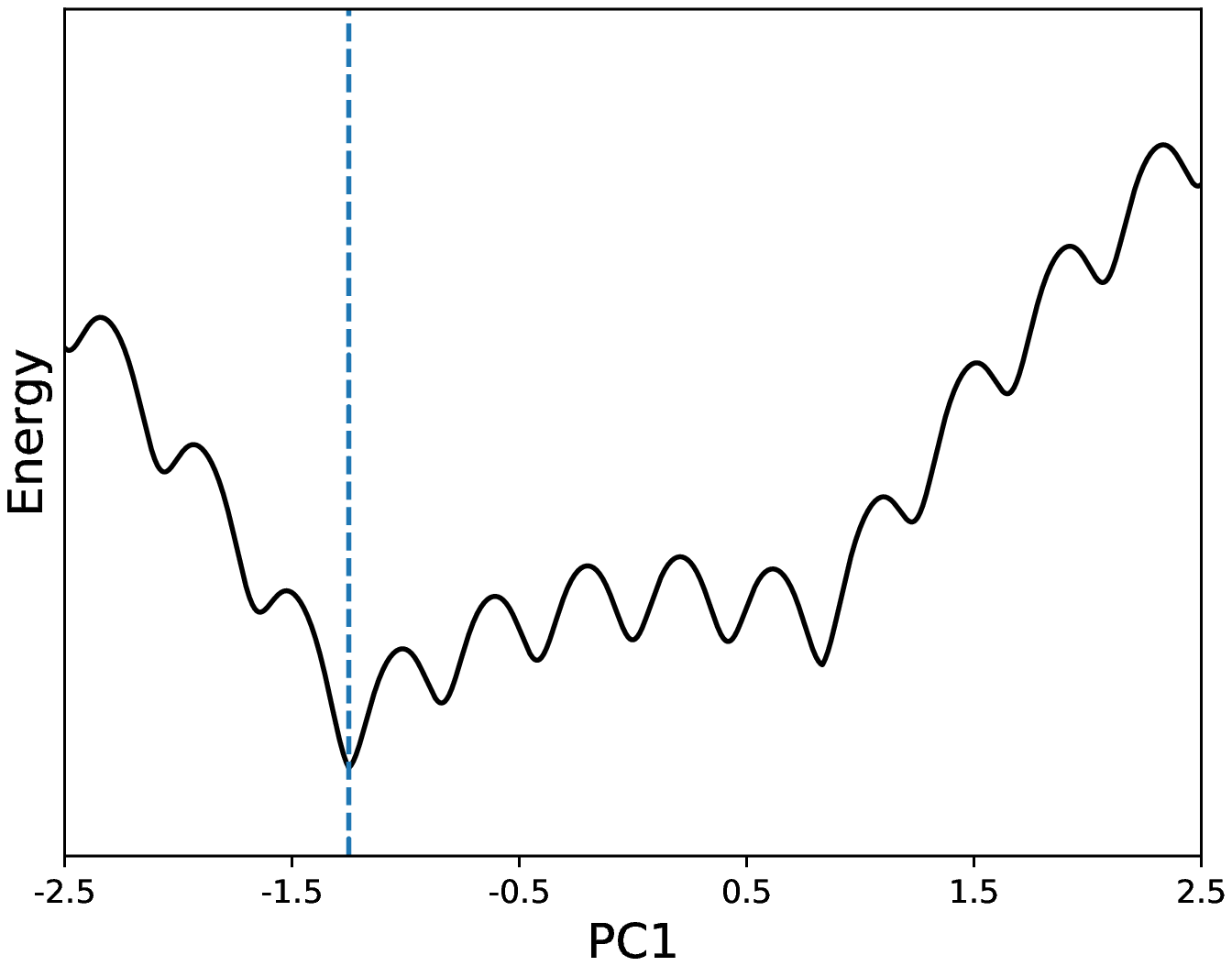}
        \caption{}
        \label{fig:gull}
    \end{subfigure}
    ~ 
    \begin{subfigure}[b]{0.23\textwidth}
        \includegraphics[width=\textwidth]{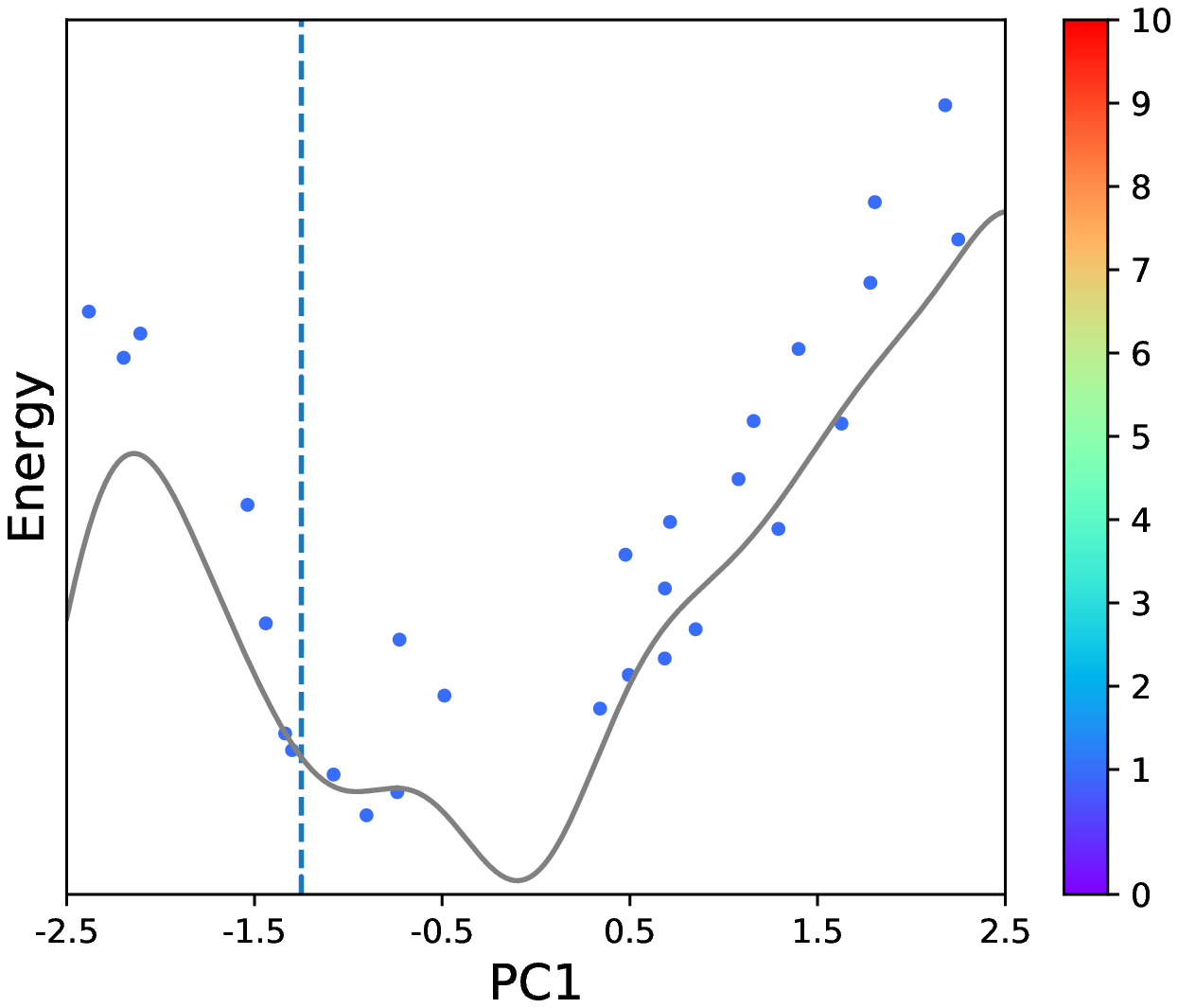}
        \caption{}
        \label{fig:tiger}
    \end{subfigure}
    ~ 
   \begin{subfigure}[b]{0.23\textwidth}
        \includegraphics[width=\textwidth]{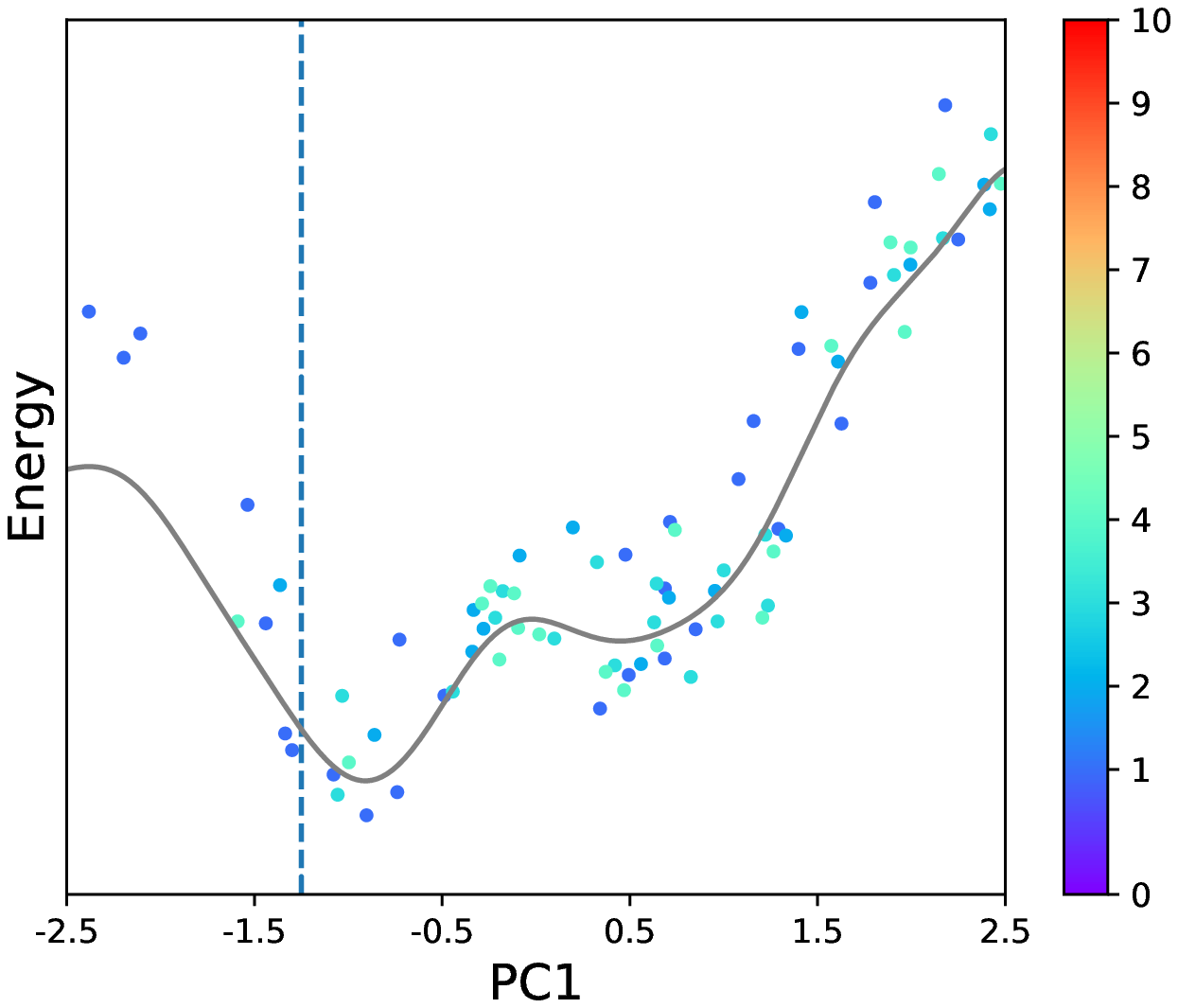}
        \caption{}
        \label{fig:gull}
    \end{subfigure}
        ~ 
   \begin{subfigure}[b]{0.23\textwidth}
        \includegraphics[width=\textwidth]{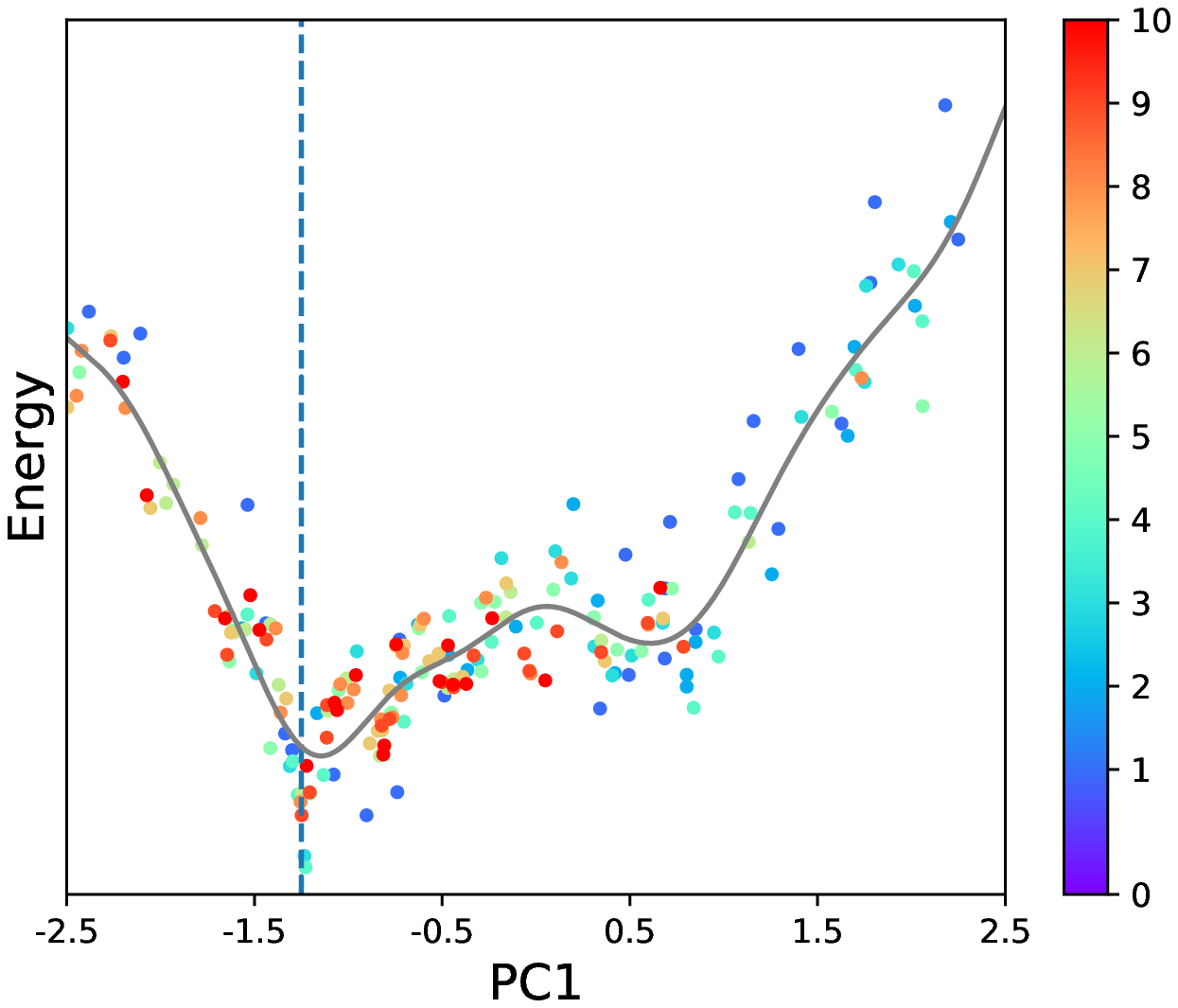}
        \caption{}
        \label{fig:gull}
    \end{subfigure}
         ~ 
   \begin{subfigure}[b]{0.23\textwidth}
        \includegraphics[width=\textwidth]{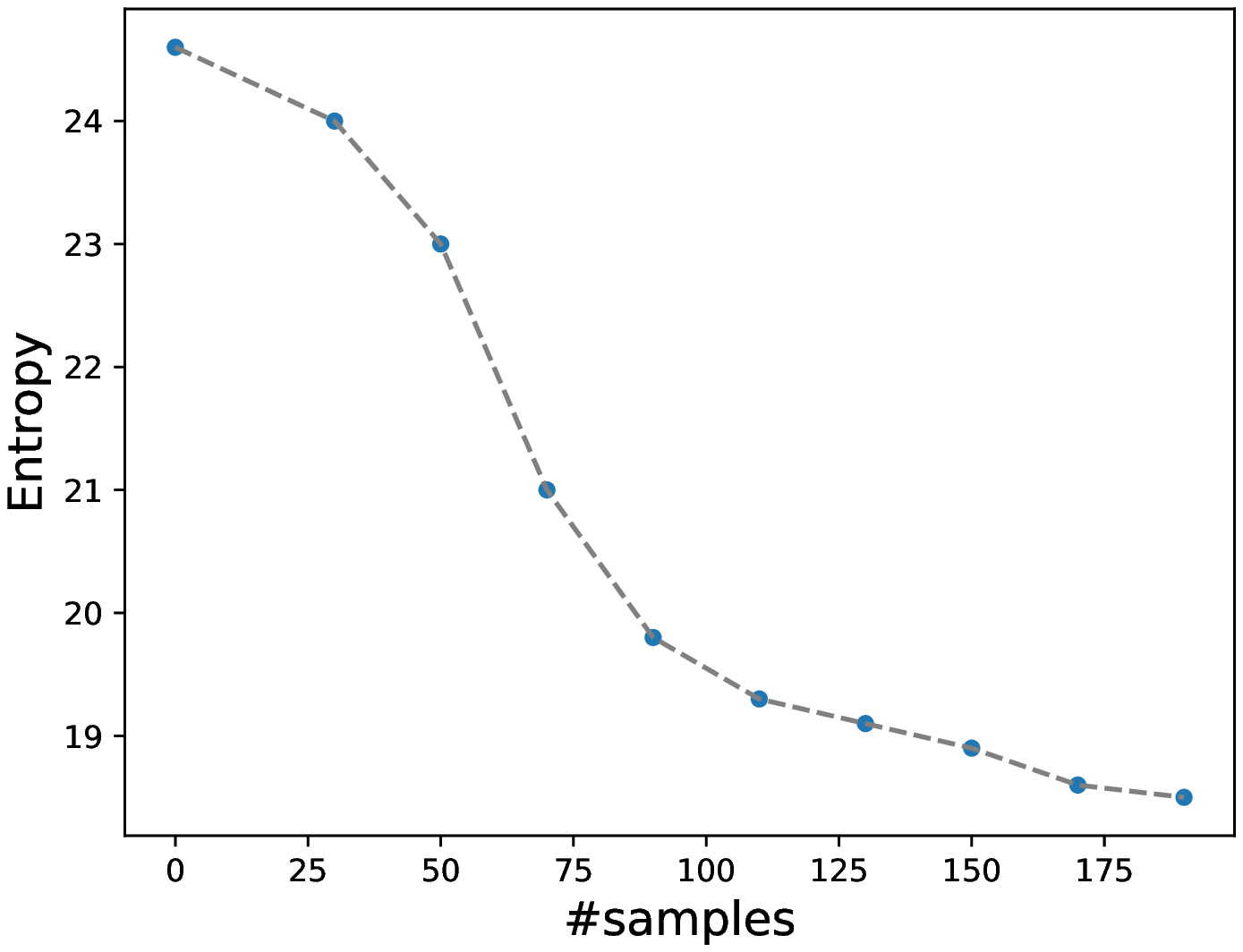}
        \caption{}
        \label{fig:gull}
    \end{subfigure}
     ~ 
   \begin{subfigure}[b]{0.23\textwidth}
        \includegraphics[width=\textwidth]{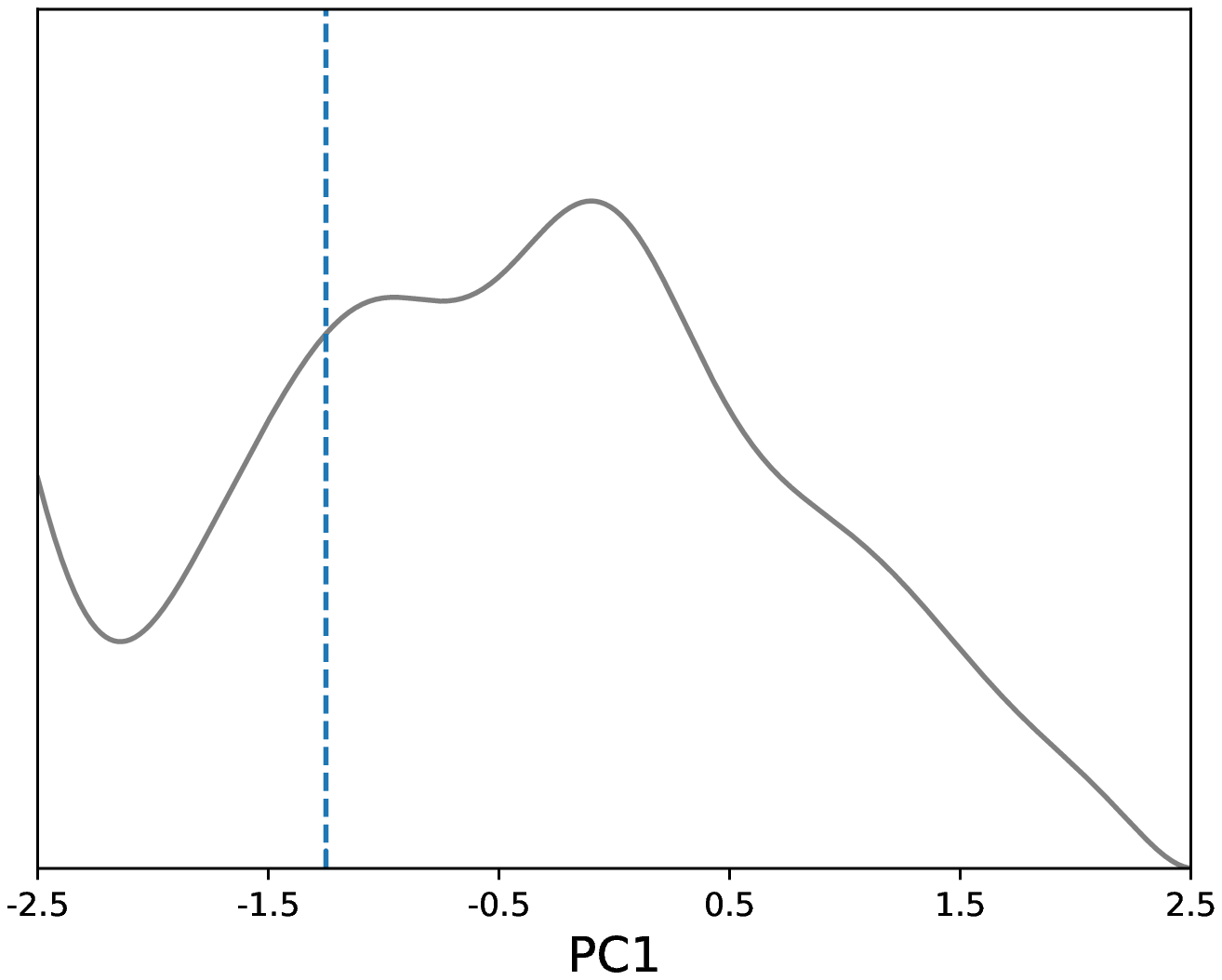}
        \caption{}
        \label{fig:gull}
    \end{subfigure}
      ~ 
   \begin{subfigure}[b]{0.23\textwidth}
        \includegraphics[width=\textwidth]{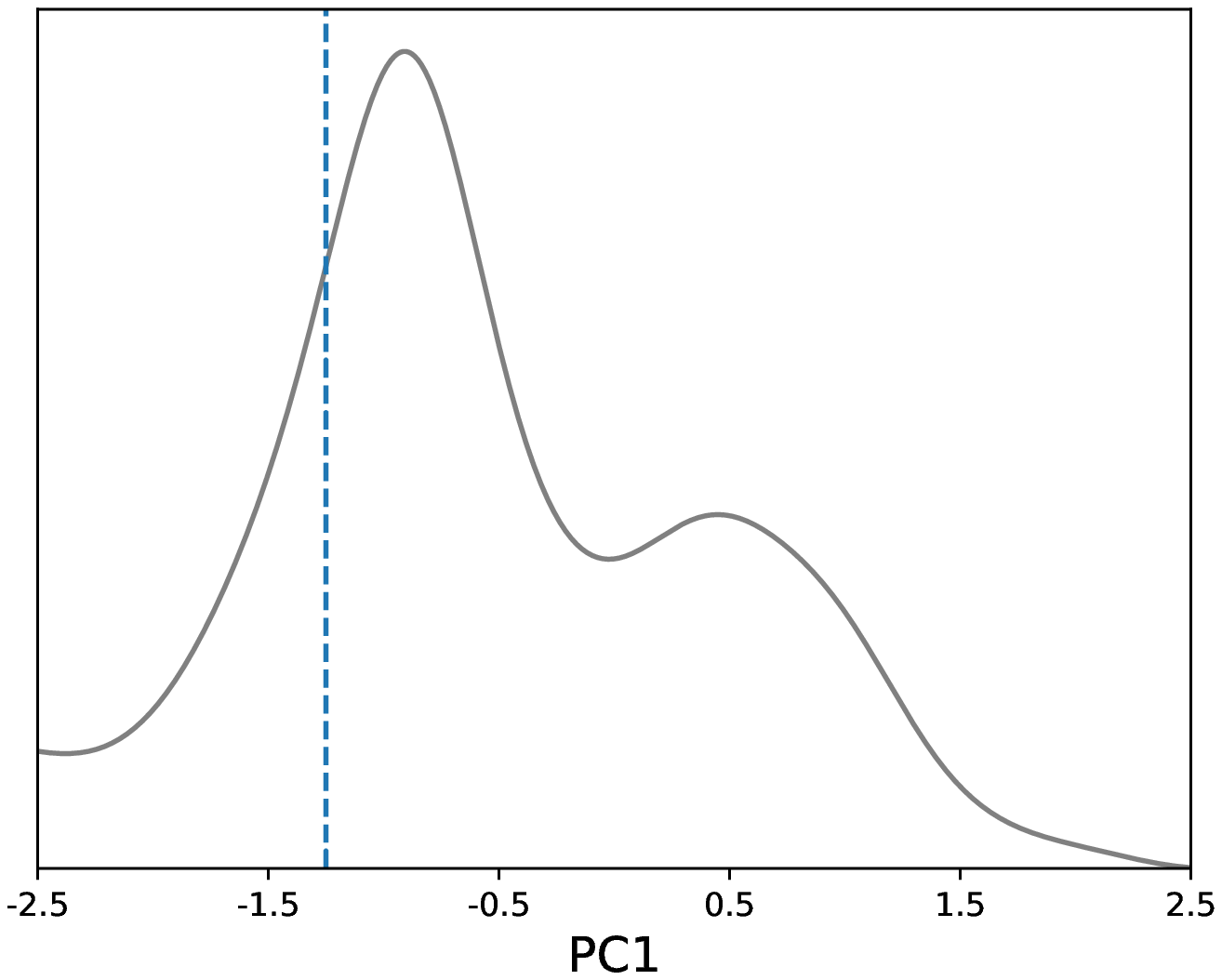}
        \caption{}
        \label{fig:gull}
    \end{subfigure}
      ~ 
   \begin{subfigure}[b]{0.23\textwidth}
        \includegraphics[width=\textwidth]{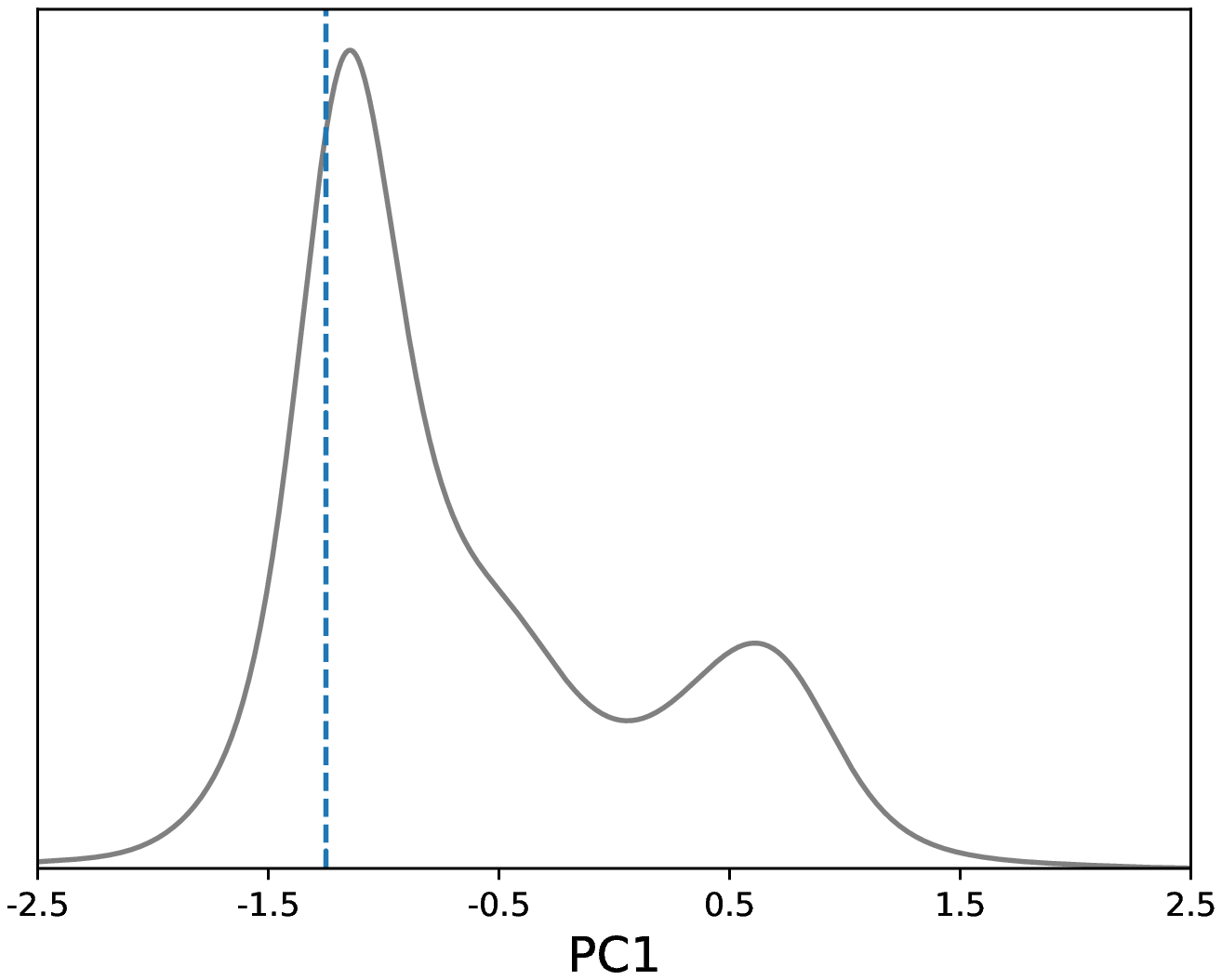}
        \caption{}
    \end{subfigure}
    \vspace{-2em}
    \caption{Illustration of the Bayesian active  learning (BAL) algorithm. (A): A typical energy landscape projected onto the first principle component (PC1). The dashed line indicates the location of the optimal solution. (B)-(D): The samples (dots) and the kriging regressors (light curves) at the stage of the 1st, 4th and 10th iteration, respectively. Samples are colored from cold to hot by the iterations and those in the same iteration have the same color. (E): The entropy (measuring uncertainty) of the posterior reduces as the number of samples increases. Its quick drop from 30 to 100 samples corresponds to a drastic change of kriging regressor, which suggests increasing exploitation in possible function basins. After 100 samples, the entropy goes down more slowly, meaning that the regressor only has a small amount of adjustment.   (F)-(H): The corresponding posterior distributions for (B)-(D), respectively.}
\label{fig:illustration}
\vspace{-2em}
\end{figure*}

\begin{methods}
\vspace{-2em}
\section{Materials and Methods}

\subsection{Mathematical Formulation}

We consider a black-box function $f(\bm{x})$ (e.g. $\Delta G$, the change in the Gibbs free energy upon protein-protein interaction) that can only be evaluated at any sample with an expensive yet noisy observation $y(\bm{x})$ (e.g. modeled energy difference).  Our goal in optimization is 
$$\vspace{-0.5em}\bm{x}^*=\arg\min_{\bm{x}\in\mathcal{X}} {f(\bm{x})}$$ 
(e.g. the native structure of a protein complex). And our goal in uncertainty quantification is the probability that $\hat{\bm{x}}$, a prediction of $\bm{x}^*$, falls in an interval $[lb, ub]$ of quality relative to $\bm{x}^*$: 
$$P\left(lb\leqslant Q(\hat{\bm{x}},\bm{x}^*) \leqslant ub \right) = 1-\sigma$$ 
where $1-\sigma$ is the confidence level; $[lb, ub]$ is the confidence interval in the solution quality; and $Q(\cdot,\cdot)$, the quality of interest measuring some distance or dissimilarity, can be an Euclidean norm (as in our assessment for test functions), another distance metric, or other choices of users (for instance, iRMSD as in our assessment for protein docking with $ub=4$\AA).

\subsection{Bayesian active learning with a posterior of $\bm{x}^*$}

We address the problem above in a Bayesian perspective: instead of treating $\bm{x}^*$ as a fixed point, we model $\bm{x}^*$ as a random variable and construct its probability distribution, $p(\bm{x}^*|\mathcal{D})$, given samples $\mathcal{D}=\{(\bm{x}, y)\}$.  This probability distribution, carrying the belief and the uncertainty on the location of $\bm{x}^*$, is a prior when $\mathcal{D}=\varnothing$ (no sample) and a posterior otherwise.  Considering the cost of function evaluation, we iteratively collect new samples in iteration $t$ (where all samples collected by the end of the $t$-th iteration are denoted $\mathcal{D}^{(t)}$) based on the latest estimated posterior, $p(\bm{x}^*|\mathcal{D}^{(t-1)})$; and we update the posterior $p(\bm{x}^*|\mathcal{D}^{(t)})$ based on $\mathcal{D}^{(t)}$.  An illustration of the iterative approach is given in Fig. \ref{fig:illustration}. 

For optimization, we set $\hat{\bm{x}}$ to be the best sample with the lowest $y$ value given a computational budget (reflected in the number of samples or iterations). For UQ, given the posterior $p(\bm{x}^*|\mathcal{D}^{(t)})$ in the final iteration, one can propagate the inferred uncertainty in $\bm{x}^*$ forwardly to that in the quality of interest, $Q(\hat{\bm{x}},\bm{x}^*)$, for the found $\hat{\bm{x}}$, using techniques such as Markov chain Monte Carlo.

\subsubsection{Non-parametric posterior of $\bm{x}^*$}

We propose to use the Boltzmann distribution to describe the posterior
\vspace{-1em}
$$p(\bm{x}^*|D^{(t)}) \propto \exp(-\rho\cdot\hat{f}(\bm{x}))$$
where $\hat{f}(\bm{x})$ is an estimator for $f(\bm{x})$, and $\rho$ is a parameter (sometimes $1/RT$ where $R$ is the gas constant and $T$ the temperature of the molecular system). 

To iteratively guide the expensive sampling and balance between exploration and exploitation in a data efficient way, we choose $\rho$ to follow an adaptive annealing schedule over iteration $t$: 
\vspace{-1em}
$$ \rho_t=\rho_0\cdot\exp((h^{(t-1)}_p)^{-1}n_t^\frac{1}{d}) $$
where $\rho_0$,  the initial $\rho$, is a parameter; $h^{(t-1)}_p)$ is the (continuous) entropy of the last-iteration posterior, a shorthand notation for $h(p(\bm{x}^*|D^{(t-1)}))$; $n_t=|\mathcal{D}^{(t)}|$ is the number of samples collected so far; and $d$ is the dimensionality of the search space $\mathcal{X}$.  

This annealing schedule is inspired by the adaptive simulated annealing (ASA) \citep{DBLP:journals/corr/cs-MS-0001018}, especially the exponential form and the $n_t^\frac{1}{d}$ term. However, we use the $(h^{(t-1)}_p)^{-1}$ term rather than a constant as in ASA so that we exploit all historic samples $\mathcal{D}^{(t)}$. In this way, as the uncertainty of $\bm{x}^*$ decreases, $\rho_t$ increases and shifts the search toward exploitation.     

The function estimator $\hat{f}(\bm{x})$ also updates iteratively according to the incrementaly increasing $n_t$ samples $\mathcal{D}^{(t)}=\{(\bm{x}_i, y_i)\}_{i=1}^{n_t}$.  We use a consistent and unbiased Kriging regressor  \citep{chiles_geostatistics:_2012} which is known to be the best unbiased linear estimator (BLUE): 
$$\vspace{-1em} \hat{f}(\bm{x})=f_0(\bm{x}) + (\bm{\kappa}(\bm{x}))^T(\bm{K}+\epsilon^2\bm{I})^{-1}(\bm{y}-\bm{f_0}) $$
where $f_0(\bm{x})$ is the prior for $E[f(\bm{x})]$; $\bm{\kappa}(\bm{x})$ is the kernel vector with the $i$th element being the kernel, a measure of similarity, between $\bm{x}$ and $\bm{x}_i$;
$\bm{K}$ is the kernel matrix with the $(i,j)$ element being the kernel between $\bm{x}_i$ and $\bm{x}_j$; $\bm{y}$ and $\bm{f_0}$ are the vector of $y_1,\ldots,y_{n_t}$ and $f_0(\bm{x}_1),\ldots,f_0(\bm{x}_{n_t})$, respectively; and $\epsilon$ reflects the noise in the observation and is estimated to be 2.1 as the prediction error for the training set. 

We derive the Kriging regressor in the Supporting Information (\textbf{SI}) Sec. 1.2.2. And we will use the regressor to evaluate binding energy and estimate iRMSD for UQ over multiple regions in Sec. 2.5.

\subsubsection{Adaptive sampling based on the latest posterior}

For a sequential sampling policy that balances exploration and exploitation during the search for the optimum, we choose Thompson sampling  \citep{russo_tutorial_2017} which samples a batch of points in the $t$-th iteration based on the latest posterior $p(\bm{x}^*|D_{t-1})$.  This seemingly simple policy has been found to be theoretically \citep{agrawal_analysis_2011} and empirically \citep{chapelle_empirical_2011} competitive compared to other updating policies such as Upper Confidence Bound \citep{shahriari_taking_2016}. In our case, it is actually straightforward to implement given the posterior on $\bm{x}^*$.  

There are multiple reasons to collect in each iteration a batch of samples rather than a single one. First, given the high dimension of the search space, it is desired to collect adequate data before updating the posterior. Second, the batch sampling weakens the correlation among samples and make them more independent, which benefits the convergence rate of the Kriging regressor. Last, parallel computing could be trivially applied for batch sampling, which would significantly improve the algorithm throughput.

Fig. \ref{fig:illustration} gives an illustration of the algorithm behavior. The initial samples drawn from a uniform distribution leads to a relatively flat posterior whose maximum is off the function optimum (Fig. \ref{fig:illustration}F). As the iteration progresses, the uncertainty about the optimum gradually reduces(Fig. \ref{fig:illustration}E)  and newer samples are increasingly focused (Fig. \ref{fig:illustration}C,D) as the posteriors are becoming narrower with peaks shifting toward the function optimum (Fig. \ref{fig:illustration}G,H).

In our docking study, $d=12$ for a homogeneous space spanned by complex normal modes (see Sec. 2.3). We construct a prior and collect 30 samples in the first iteration and 20 in each of the subsequent iterations. We limit the number of iterations (samples) to be 31 (630) as a way to impose a computational budget.  

\subsubsection{Kernel with customized distance metric}

The kernel in the Kriging regressor for the posterior is a measure of similarity. For test functions defined in an Euclidean space, we use the radial basis function (RBF) kernel:
$$ \vspace{-2em} \kappa(\bm{x}_i,\bm{x}_j)=exp(-\frac{||\bm{x}_i-\bm{x}_j||^2}{2l^2}) $$
where $||\bm{x}_1-\bm{x}_2||^2$, a measure of dissimilarity, is the Euclidean distance and $l$,  the bandwidth of the kernel, is set as $l=l_0 \cdot n_t^{\frac{1}{d}}$ following  \citep{gyorfi_distribution-free_2002}. $l_0$, dependent on search space, is set at 2.0 for docking without particular optimization.

For protein docking, we replace the Euclidean distance in the RBF kernel with the interface RMSD (iRMSD) between two sample structures. iRMSD captures sample dissimlarity relevant to function-value dissimilarity and is independent of search-space parameterization.  

For this purpose, we also have to address two technical issues. First, protein interface information is determined by $\bm{x}^*$ and thus unknown. We instead use the putative interface seen in the samples (see more details in the \textbf{SI} Sec. 2.4). Second, kernel calculation with iRMSD is time consuming. The time complexity of iRMSD calculation is $O(N)$ and that of regressor update is $O(Nn^2)$, where $N$, the number of interfacial atoms, can easily reach hundreds or thousands, and $n$, the number of samples, can also be large. To save computing time, we develop a fast RMSD calculation method  that reduces its time complexity from $O(N)$ down to $O(1)$  (see details in \textbf{SI} Sec. 2.1).

\subsubsection{Related methods}

Current Bayesian optimization methods typically model the posterior distribution of $f(\bm{x})$ rather than that of  $\bm{x}^*=\arg\min_{\bm{x}\in\mathcal{X}} f(\bm{x})$ directly.  After modeling the posterior distribution over the functional space (a common non-parametric way is through Gaussian processes), they would subsequently sample the functional space and optimize sample functions. For instance, \cite{villemonteix_informational_2006} used Monte Carlo sampling; and  \cite{henrandez-lobato_predictive_2014} discretized the functional space to approximate the sample paths of Gaussian processes using a finite number of basis functions then optimized each sample path to get one sample of $\bm{x}^*$.  The two-step approach of current Bayesian optimization methods involve intensive sampling and (non-convex) optimization that is computationally intensive and not pragmatically useful for protein docking.

To our knowledge, \citeauthor{ortega_nonparametric_2012} presented the only other study that directly models the posterior distribution over the optimum \citep{ortega_nonparametric_2012}.  Both methods fall in the general category of Bayesian optimization and use consistent non-parametric regressors.  However, we prove in Sec. 1.2 of the \textbf{SI} that their regressor is biased whereas our Kriging regressor is unbiased.  We explain in Sec. 1.1 there that their annealing schedule (temperature control) only considers the pairwise distance between samples without location awareness and is independent of dimensionality $d$; whereas ours has a term involving location-aware global uncertainty and generalizes  well to various dimensions. Beyond those theoretical comparison, we also included empirical results to show the superior optimization and UQ performances for BAL.

The rest of the Materials and Methods section involve methods specific to the protein docking problem:  parameterization, dimensionality reduction, and range reduction of the search space $\mathcal{X}$;  machine learning model as $y(\bm{x})$, i.e., an energy model for encounter complexes; uncertainty quantification for a predicted structure or a list of predictions as well as the corresponding quality estimation for ranking predictions or classifying their nativeness. 

\vspace{-1.5em}
\subsection{Conformational Sampling in $\mathcal{X}$}

In protein docking the full search space $\mathcal{X}$ captures the degrees of freedom for all atoms involved. Let one protein be receptor whose position is fixed and the other be ligand. And let $N_\text{R}$, $N_\text{L}$, and $N$ be the number of atoms for the receptor, the ligand, and the complex respectively.  Then $\mathcal{X}=\mathbb{R}^{3N-6}$ is a Euclidean space whose dimension easily reaches $10^4$ for a small protein complex without surrounding solvent molecules. If accuracy is sacrificed for speed, proteins can be (unrealistically) considered rigid and $\mathcal{X}= SE(3)=\mathbb{R}^3\times SO(3)$ (ligand translations and rotations) is a Riemannian manifold. Docking methods fall in the spectrum between these two ends that are represented by all-atom molecular dynamics and FFT rigid docking, respectively.  For instance, one can consider locally rigid pieces of a protein rather than a globally rigid protein, then $\mathcal{X}$ becomes the product of many $SE(3)$ for local rigidity \citep{mirzaei2015energy}; or one can model individual proteins' internal flexible-body motions normal modes on top of the ligand rigid-body motions, thus $\mathcal{X}$ becomes the product of $\mathbb{R}^K$ (where $K\ll N_\text{R/L}$) and $SE(3)$ \citep{moal_swarmdock_2010}.

From the perspective of optimization and UQ, both the high-dimensionality of $\mathbb{R}^{3N-6}$ and the geometry of the lower-dimensional manifold present challenges. Almost all dimensionality reduction efforts in protein docking impose conditions (such as aforementioned local or global rigidity) in the full Euclidean space and lead to embedded manifolds difficult to (globally) optimize over. The challenge from the manifold has been either disregarded in protein docking or addressed by the local tangent space \citep{shen2007optimizing,shen2008protein,mirzaei2015energy}.  

Could and how could the dimensionality of the conformational space be reduced while its geometry maintains homogeneity and isotropy of a Euclidean space and its basis vectors span conformational changes of proteins upon interactions? In this subsection we give a novel approach to answer this question for the first time. 
In contrast to common conformational sampling that separates internal flexible-body motions (often Euclidean) and external rigid-body motions (a manifold) \citep{marze_efficient_2018}, we re-parameterize the space into a Euclidean space spanned by complex normal modes \cite{oliwa_cnma:_2015} blending both flexible- and rigid-body motions. The mapping preserves distance metric in the original full space.  We further reduce the dimensionality and the range in the resulting space.  

\subsubsection{Complex normal modes blend flexible- and rigid-body motions}

We previously introduced complex normal mode analysis, cNMA \citep{oliwa_cnma:_2015}, to model conformational changes of proteins during interactions. Using encounter complexes from rigid docking, cNMA extends anisotropic network model (ANM) to capture both conformational selection and induced fit effects. After the Hessian matrix is projected to remove the rigid-body motion of the receptor, its non-trivial eigenvectors $\pmb{\mu}_j$ ($j=1,\ldots,3N-6$) form orthonormal basis vectors. We showed that $\pmb{\mu}_j^\text{R}$, the components of the complex normal modes, better capture the direction of individual proteins' conformational changes than conventional NMA did \citep{oliwa_cnma:_2015}. We also showed that the re-scaled eigenvalues for these components, $ \lambda_j^\text{R}=\frac{\lambda_j}{||\pmb{\mu}_j^\text{R}||^2}$, can be used to construct features for machine learning and predict the extent of the conformational changes. 

\subsubsection{Dimensionality reduction}
In this study we focus on the motions of a whole complex rather than individual proteins and develop sampling techniques for protein docking. Each single complex normal mode simultaneously captures concerted flexible-body motions of individual proteins (receptor and ligand) and rigid-body motion of the non-fixed protein (ligand).  They together span a homgeneous and isotropic Euclidean space where the distance between two points is exactly the RMSD between corresponding complex structures.  

For dimensionality reduction in the resulting space, we choose the first $K_1$ non-trivial eigenvectors $\pmb{\mu}_j$ ranked by increasing eigenvalues $\lambda_j$; and we additionally include $K_2$ $\pmb{\mu}_j$ (not in the first $K_1$) ranked by increasing $\lambda_j^R$.  In other words, we choose $K_1$ slowest modes for the complex and another $K_2$ for the receptor as the set of basis vectors $\mathcal{B}$. $K_1$ and $K_2$ are set at 9 and 3, respectively, leading to the dimension  of the reduced space to $d=12$. A supplemental video illustrates the motion of the slowest such modes.

\subsubsection{Range reduction}

For range reduction in the dimension-reduced space, we perturb a starting complex structure $\vec{C}\in\mathbb{R}^{3N-6}$ along aforementioned basis vectors to generate sample $\vec{C}_0$ while enforcing a prior on the scaling factor $s$ in the first iteration. Specifically  
$$ \vec{C} = \vec{C}_0 + \sum_{j \in \mathcal{B}} r_{j}\frac{s}{\sqrt{\lambda_j}}\cdot\pmb{\mu}_j $$
where $r_{j}$, the coefficient of the $j$th normal mode $\pmb{\mu}_j$, is uniformly sampled on $S^d$, the surface of a $d$-dimensional standard sphere with a unit radius. The scaling factor $s$ is given by
$$ s=\frac{\tau_\text{R}}{{\frac{1}{\sqrt{N_\text{R}}}{\norm{\sum\limits_{j\in \mathcal{B}}\frac{r_j}{\sqrt{\lambda_j}}\cdot \pmb{\mu}_j^\text{R}}}}} $$

where $\tau_\text{R}$ is the estimated conformational change (measured by RMSD) between the unbound and the bound receptor. Note that $\pmb\mu_i^R$'s are not orthonormal to each other. 

We previously predicted $\tau_\text{R}$ by a machine learning model giving $\widehat{\text{RMSD}}_\text{R}$, a single value for each receptor  \citep{chen_predicting_2017}. Here we replace $\widehat{\text{RMSD}}_\text{R}$ with a predicted  distribution by multiplying it to a truncated normal distribution $N(\mu=0.99, \sigma^2=0.096)$ within [0, 2.5].  The latter distribution is derived by fitting the ratios between the actual and the predicted values, $\text{RMSD}_\text{R}/\widehat{\text{RMSD}}_\text{R}$, for 50 training protein complexes (see more details about datasets in Sec. 2.7 and those about distribution fitting in Sec. 2.2 of the \textbf{SI}). Therefore, our parameterization produces $\bm{x}=s\cdot\bm{r}\in\mathbb{R}^d$ whose prior is derived as above.

Since the ligand component of complex normal modes include simultaneous flexible- and rigid-body motions, conformational sampling could lead to severely distorted ligand geometry.  We thus further restrict the ligand perturbation $\Delta_\text{L}$ (flexible- and rigid-body together) to be within $\overline{\Delta}_\text{L}$  
$$
\Delta_\text{L}=\sqrt{\frac{1}{N_\text{L}}}\norm{\sum_{j\in \mathcal{B}}r_j\frac{s}{\sqrt{\lambda_j}}\cdot \pmb\mu_j^\text{L}} \leqslant  \overline{\Delta}_\text{L}
$$
We set $\overline{\Delta}_\text{L}$ at 6\AA\ according to the average size of binding energy attraction basins seen in conformational clusters \citep{kozakov2005optimal}. For samples generated from the aforementioned prior or the updated posterior, we reject those violating the ligand perturbation limit.  We discuss about the feasibility of the search region in \textbf{SI} Sec. 2.3.

\subsection{Energy Model $y(\bm{x})$ for Sampling}
\subsubsection{Binding affinity prediction for sampled encounter complexes}

We introduce a new energy model based on binding affinities $K_d^\prime(\bm{x})$ of structure samples $\bm{x}$ that are often encounter complexes. The model assumes that $K_d^\prime$ correlates with $K_d$, the binding affinity of the native complex,  and deteriorates with the increase of the sample's iRMSD (the encounter complex is less native-like):  

$$K_{d}^\prime(\bm{x}) = K_d\cdot\exp(\alpha \cdot ({\text{iRMSD}(\bm{x})})^q)$$

where $\alpha$ and $q$ are hyper-parameters optimized through cross-validation. In other words, we assume that the fraction of binding affinity loss is exponential in a polynomial of iRMSD. Therefore, the binding energy, a machine learning model $y(\bm{x};\bm{w})$ of parameters $\bm{w}$ can be represented as
\vspace{-2em}
$$y(\bm{x};\bm{w})=RT\ln(K_d^\prime(\bm{x}))=RT\ln(K_d)+RT\alpha \cdot ({\text{iRMSD}(\bm{x})})^q$$ 
\vspace{-0.5em}
Given an observed or regressed $y(\bm{x})$ value, one can estimate $\text{iRMSD}(\bm{x})$ with given $K_d$.

\subsubsection{Machine learning }

We train machine learning models, including ridge regression with linear or RBF kernel and random forest, for $y(\bm{x};\bm{w})$. The 8 features include changes upon protein interaction in energy terms such as internal energies in bond, angle, dihedral and Urey-Bradley terms, van der Waals, non-polar contribution of the solvation energy based on solvent-accessible surface area (SASA), and electrostatics modeled by Generalized Born with a simple SWitching (GBSW), all of which are calculated in a CHARMM27 force field. 

We use the same training set of 50 protein pairs (see details in \textbf{SI} Sec. 2.5) as in predicting the extent of conformational change. From rigid docking and conformational sampling we generate 13,004 complex samples for 50 protein pairs, including 6,464 near-native and 6,540 non-native   examples in the training set. Hyper-parameters of ridge regression with RBF kernel as well as random forest are optimized by cross-validation.  And model parameters $\bm{w}$ are trained again over the entire training set with the best hyper-parameters.  More details can be found in Sec. 2.6 of the \textbf{SI}.  For the assessment, we use the test set \textbf{a} of 26 protein pairs (again in \textbf{SI} Sec. 2.5) and generate 20 samples similarly for each of the 10 initial docking results for each pair, leading to 5,200 cases.

\subsection{Uncertainty Quantification for Protein Docking}

A unique challenge to protein docking refinement is that, instead of optimization and UQ in a single region $\mathcal{X}$, we may do so in $K$ separate ones $\mathcal{X}_i$ ($i=1,\ldots,K$) where each $\mathcal{X}_i$ is a promising conformational region/cluster represented by a initial docking result. This is often necessitated by the fact that the extremely rugged energy landscape is populated with low-energy basins separated by frequent high-energy peaks in a high-dimensional space, thus preferably searched over multiple stages \citep{vajda2009convergence}.

Our goal of UQ for protein docking results is to determine, for each $\hat{\bm{x}_i}$ -- the prediction in $\mathcal{X}_i$ (the $i$th structure model), its  quality bounds [$lb$, $ub$] such that 
$$P\left(lb\leqslant Q(\hat{\bm{x}_i},\bm{x}^*) \leqslant ub \right) = 1-\sigma$$ 
where the quality of interest $Q(\hat{\bm{x}},\bm{x}^*)$ here is iRMSD between a predicted and the native structure and $1-\sigma$ a desired confidence level. 

To that end, we forwardly propagate the uncertainty from $\bm{x}^*$ (native structure) to iRMSD, given the final posterior $p(\bm{x}^*|\mathcal{D}_t)$ in individual regions (local posteriors).  Specifically, we generate 1,000,000 samples following the local posterior using Markov chain Monte Carlo, evaluate their binding energies using the Kriging regressor, and estimate their iRMSD using our binding affinity prediction formula inversely.  We then use these sample iRMSD values to determine confidence intervals [$lb$, $ub$] for various confidence score $1-\sigma$ so that $P(\text{iRMSD}<lb)=P(\text{iRMSD}>ub)=\sigma/2$.

\subsection{Confidence scores for near-nativeness}

We next calculate the probability that a prediction $\hat{\bm{x}}_i$ is near-native, i.e., $P(Q(\hat{\bm{x}}_i,\bm{x}^*)\leqslant 4\AA)$ \citep{mendez_assessment_2005}. Calculating this quantity would demand the probability that the native structure lies in the $i$th conformational region / cluster, $P(\bm{x}^* \in \mathcal{X}_i)$ ($P(\mathcal{X}_i)$ in short) as well as that the probability that it lies in all the $K$ regions, $P(\bm{x}^* \in \cup_{i=1}^{K}\mathcal{X}_i)$ ($P(U_K)$ in short).  By following the chain rule we easily reach 
\vspace{-1em}
$$
\begin{array}{rl}
& P(\text{iRMSD}(\hat{\bm{x}}_i,\bm{x}^*)\leqslant 4)= \\ & P(\text{iRMSD}(\hat{\bm{x}}_i,\bm{x}^*)\leqslant 4 |\mathcal{X}_i)\cdot P(\mathcal{X}_i|U_K)\cdot P(U_K)
\end{array}
$$. 
Here we use the fact that $\{\bm{x}^* \in \mathcal{X}_i\}\subset \{\bm{x}^* \in \cup_{i=1}^K \mathcal{X}_i\} $ and assume that $\{\text{iRMSD}(\hat{\bm{x}}_i,\bm{x}^*)\leqslant 4\} \subseteq \{\bm{x}^* \in \mathcal{X}_i\}$ (the range of conformational clusters in iRMSD is usually wider than 4\AA). 

We discuss how to calculate each of the three terms for the product.

\subsubsection{$P(\text{iRMSD}(\hat{\bm{x}}_i,\bm{x}^*)\leqslant 4 |\mathcal{X}_i)$}

If the native structure $\bm{x}^*$ (unknown) is contained in the $i$th region/cluster $\mathcal{X}_i$, what is the chance that the predicted structure $\hat{\bm{x}}_i$ (known) is within 4\AA?  We again use forward uncertainty propagation starting with the local posterior $p(\bm{x}^*|\mathcal{D}_t)$ in $\mathcal{X}_i$.  We sample 100,000 structures following the posterior with Markov chain Monte Carlo, calculate their iRMSD to the prediction $\hat{\bm{x}}_i$, and empirically determine the portion within 4\AA for the probability of interest here.  Notice that the native interface is unknown thus the putative interface is used instead. 

\subsubsection{$P(\mathcal{X}_i|U_K)$}

If the native structure is contained in at least one of the $K$ regions, what is the chance that it is in $\mathcal{X}_i$?  Following statistical mechanics, we reach 
\vspace{-1em}
$$P(\mathcal{X}_i|U_K)=\frac{Z_i}{Z} = \frac{\int_{\bm{x} \in \mathcal{X}_i}\exp(-\frac{1}{RT}\hat{f}_i(\bm{x}))d\bm{x}}{\sum_{j}^{K}\int_{\bm{x} \in \mathcal{X}_j}\exp(- \frac{1}{RT}\hat{f}_j(\bm{x}))d\bm{x}} $$ 
where $Z_i$ and $Z$ are local and global partition functions, respectively; and $\hat{f}_i(\bm{x})$ is the final Kriging regressor.  Different regions are assumed to be mutually exclusive. The integrals are calculated by Monte Carlo sampling. 

Anther approach is 
to replace $\frac{1}{RT}$ above with $\rho_i$, the final $\rho$ for temperature control in $\mathcal{X}_i$. In practice we did not find significant performance difference between the two approaches, partly due to the fact that final $\rho_i$ in various clusters / regions reached similar values for the same protein complex.

\subsubsection{$P(U_K)$}

What is chance that the native structure is within the union of the initial regions, i.e., some initial region is near-native?  The way to calculate $P(U)$ is very similar to that in uncertainty quantification. Specifically, 100,000 structures are sampled following the posterior of each region $\mathcal{X}_i$, evaluated for binding energy using the Kriging regressor $\hat{f}_i(\bm{x})$, and estimated with iRMSD using the binding affinity predictor formula inversely.  We empirically calculate the portion $q_i$ in which sample iRMSD values are above 4\AA. Assuming the independence among regions, we reach 
$P(U_K)=1 - \prod_{i=1}^K q_i$.

\subsection{Data sets}

We use a comprehensive protein docking benchmark set 4.0 \cite{hwang_protein-protein_2010} of 176 protein pairs that diversely and representatively cover sequence and structure space, interaction types, and docking difficulty. We split them into a training set, test sets \textbf{a} and \textbf{b} with stratified sampling to preserve the composition of difficulty levels in each set.  The ``training" set is not used for tuning BAL parameters (Sec. 2.2).  Rather, it is just for training energy model  $(y(\bm{x};\bm{w})$ in Sec. 2.4) and conformational-change extent prediction ($\tau_R$ in Sec. 2.3.3). The training and test \textbf{a} sets contain 50 and 26 pairs with known $K_d$ values (\citep{kastritis_are_2010}), respectively. And the test set \textbf{b} contain 100 pairs with $K_d$ values predicted from sequence alone \citep{yugandhar2014protein}.

We also use a smaller yet more challenging CAPRI set of 15 recent CAPRI targets  \citep{chen_predicting_2017}.  Unlike the benchmark set for unbound docking, the CAPRI set contains 11 cases of homology docking, 8 of which start with just sequences for both  proteins and demand homology models of structures before protein docking.  Compared to the benchmark test set of 86 (68\%), 22 (18\%) and 18 (14\%) cases classified rigid, medium, and flexible, respectively; the corresponding statistics for the CAPRI set are 4 (27\%), 5 (33\%) and 6 (40\%), respectively. Their $K_d$ values are also predicted from sequence alone. 

The complete lists of the benchmark sets and the CAPRI set, with difficulty classification, are provided in Sec. 2.5 of the \textbf{SI}.  

For each protein pair, we use 10 distinct encounter complexes as starting structures ($K=10$). As reported previously \citep{chen_predicting_2017}, those for the benchmark sets are top-10 cluster representatives by ZDOCK , kindly provided by the Weng group; and those for the CAPRI set are top-10 models generated by the ZDOCK webserver. 

\end{methods}

\section{Results and Discussion}
\subsection{Optimization and UQ Performance on Test Functions}

We first tested our BAL algorithm on four non-convex test functions of various dimensions and compared it to particle swarm optimization (PSO) \citep{kennedy1995particle,clerc2002particle}, an advanced optimization algorithm behind a very  successful protein-docking method SwarmDock \citep{moal_swarmdock_2010}. Detailed settings are provided in Sec. 2.7 of the \textbf{SI}.

For optimization we assess  $||\hat{\bm{x}}-\bm{x}^*||$, the distance between the predicted and actual global optima, a measure of direct relevance to the quality of interest in protein docking -- iRMSD.  
Compared to PSO, BAL made predictions that are, on average, closer to the global optima with smaller standard deviations; and the improvement margins increased with the increasing dimensions (Table S8 in the Supplementary Material).

For UQ we assess $r_\text{90}$, the distance upper bound of 90\% confidence, i.e., $P(||\hat{\bm{x}}-\bm{x}^*||\leqslant r_\text{90})=1-\sigma=90\%$. The metrics to assess $r_\text{90}$ include $\eta$, the relative error ($\eta=|\frac{r_\text{90}}{||\hat{\bm{x}}-\bm{x}^*||}-1|$); and $\hat{P}$, the portion of the confidence intervals from 100 runs that actually encompass the corresponding global optimum. We found in Table S9  
that our confidence intervals are usually tight judging from $\eta$ and they contain the global optima with portions $\hat{P}$ close to 90\%, the desired confidence level. The portions agreed less with the desired confidence level for some functions as the dimensionality increase, which suggests the challenge of optimization and UQ in higher dimensions.

\subsection{Evaluation of Our Energy Models}

\begin{figure*}[!htb]
    \centering
    \begin{subfigure}[b]{0.25\textwidth}
     \includegraphics[width=\textwidth]{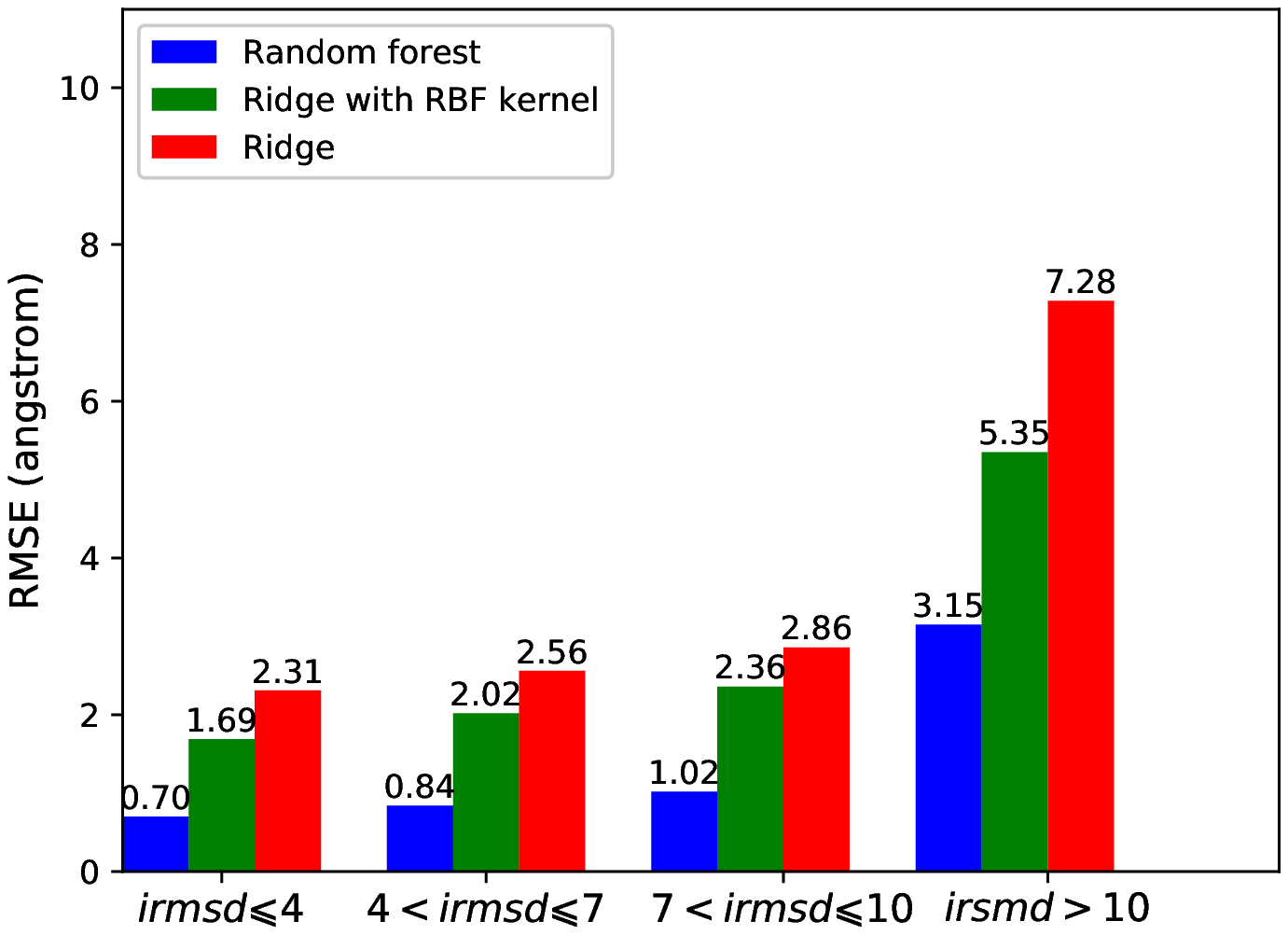}
        \caption{}
        \label{fig:score-iRMSD-train}
    \end{subfigure}
    ~ 
    \begin{subfigure}[b]{0.25\textwidth}
       \includegraphics[width=\textwidth]{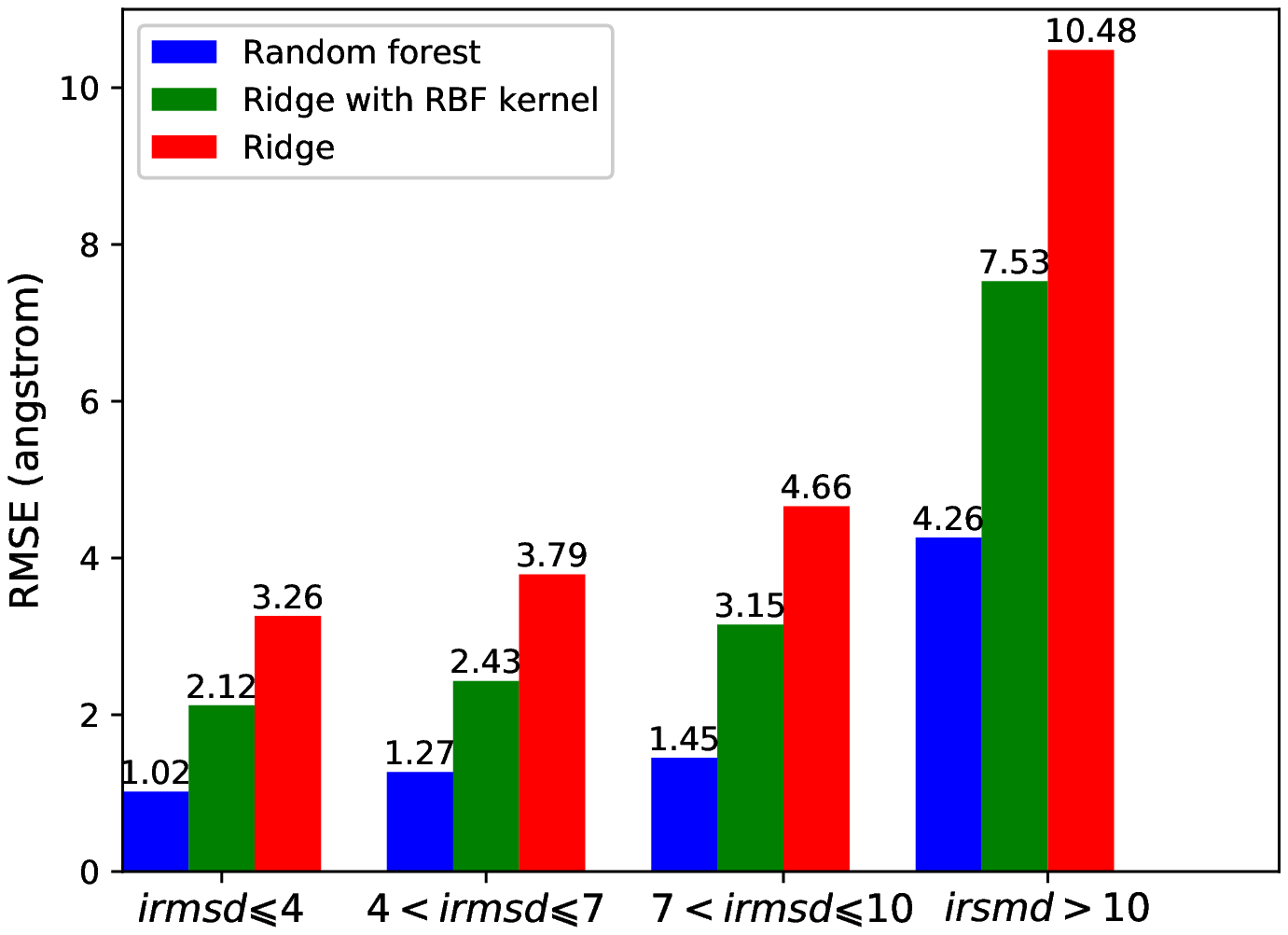}
        \caption{}
        \label{fig:score-iRMSD-test}
    \end{subfigure}
    ~ 
    \begin{subfigure}[b]{0.25\textwidth}
    \includegraphics[width=\textwidth]{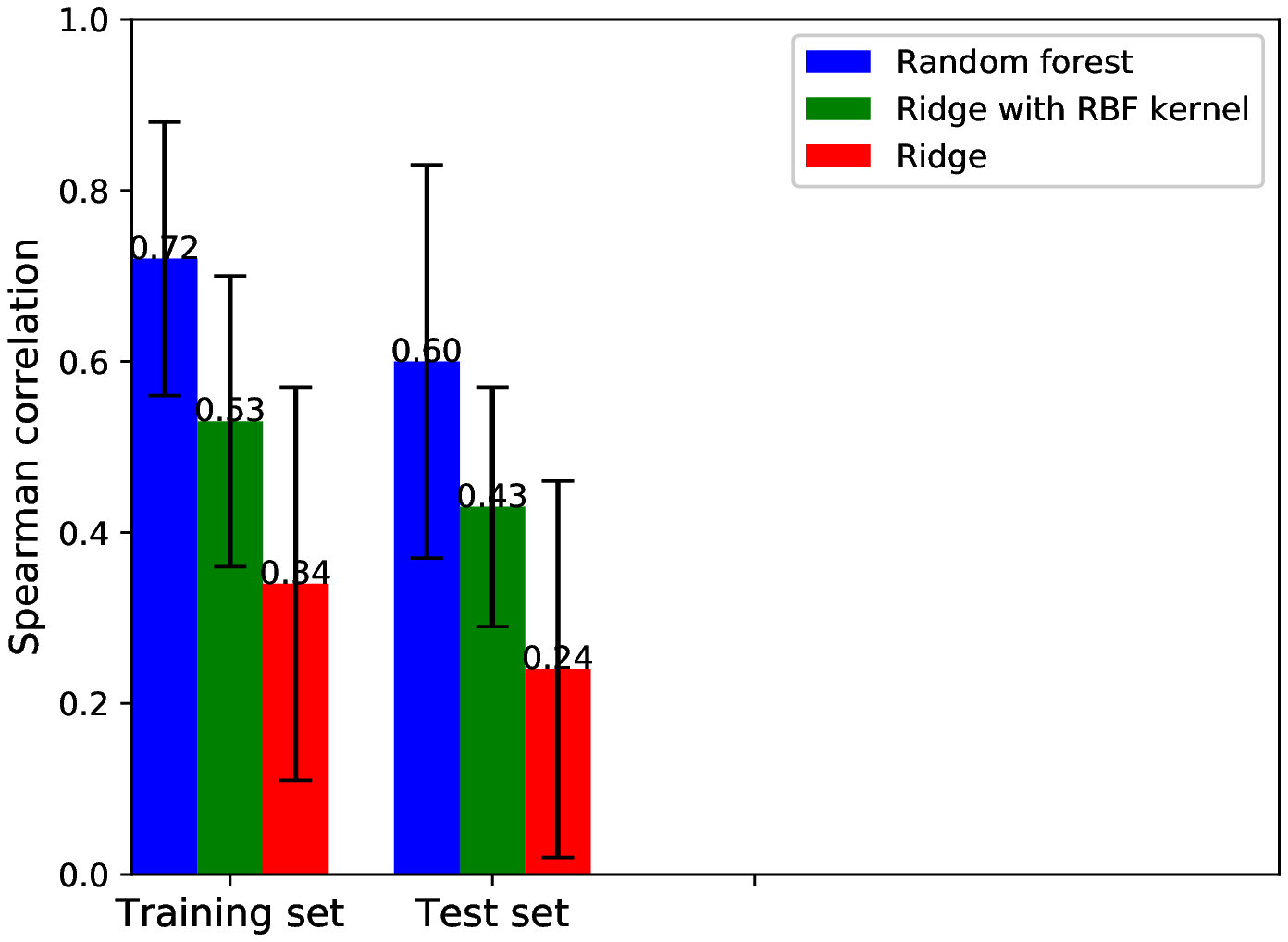}
        \caption{}
        \label{fig:score-dG-native-Spearman}
    \end{subfigure}
    \vspace{-2em}
    \caption{The root mean square error (RMSE) between the predicted and actual iRMSD for (A): Training set and (B): Test set.  (C): The Spearman's $\rho$ between predicted $y(\bm{x})$ and the real iRMSD for the training and test sets. }
    \label{fig:scoringfunction}
    \vspace{-2em}
\end{figure*}

We compare the performances of three machine learning models over the training and test sets for energy model (Sec. 2.4.2).  As no actual binding affinities of encounter complexes are available, we estimated the iRMSD values based on the random forest model's binding energy prediction (Sec. 2.4.1) and compared them to the actual iRMSD (of native interfaces) using RMSE for absolute error.  Random forest gave the best performances thus used as the energy model $y(\bm{x})$ hereinafter. Specifically, performances are split to encounter complexes of varying quality (iRMSD) in Fig.\ref{fig:score-iRMSD-train} and Fig. \ref{fig:score-iRMSD-test}.  The random forest model (blue bars) led to RMSE of 0.70\AA\ (1.0\AA) for the near-native samples in the training (test) set.  The RMSEs increased slowly as iRMSD $\leqslant 10$\AA\ and did sharply beyond (a region too far from the native for refinement).  

We also assess how ``funnel-like" the energy model is.  We thus calculated for each protein pair the Spearman's ranking coefficient $\rho$ between the energy model and the actual iRMSD. The random forest of MM-GBSW features showed the highest $\rho$ of 0.72 and 0.60 for the training and the test sets, respectively (Fig. \ref{fig:score-dG-native-Spearman}), albeit with large deviation across protein pairs.

We lastly assess the energy model's ability to rank across protein pairs. Specifically, we estimated each native protein-complex's binding energy by setting iRMSD to be zero in the energy model and compared the estimated and actual binding energy using RMSE and Pearson's $r$ in the supplemental Table S11. The random-forest energy model achieved 2.45 (4.78) Kcal/mol in RMSE and Pearson's $r$ of 0.79 (0.75) in binding energy $\Delta G=-RTln(K_d)$ for the training (test) set.



\vspace{-2em}

\subsection{Docking Performance: Optimization}

We show the improvements in PSO and BAL solution quality (measured by the decrease of iRMSD) against the starting ZDOCK solutions in Fig. S4 of the Supplemental Material.  Speaking of the amount of improvement, BAL improved iRMSD by 1.2\AA, 0.74\AA, and 0.76\AA{} for the training, test, and CAPRI sets, respectively, outperforming PSO's corresponding measures of 0.82\AA, 0.45\AA, and 0.49\AA.  It also outperformed PSO for the more challenging near-native cases (note that BAL's iRMSD improvement for the near-native test set or CAPRI set was almost neutral).  Speaking of the portion with improvement, BAL improved iRMSD in the near-native cases for 75\%, 68\%, and 73\% of the training, test, and CAPRI sets, respectively; whereas the corresponding statistics for PSO were 59\%, 50\%, and 53\%, respectively. More split statistics based on docking difficulty can be found in Fig. S5.

\begin{figure*}[!htb]
    \centering
    \begin{subfigure}[b]{0.25\textwidth}
        \includegraphics[width=\textwidth]{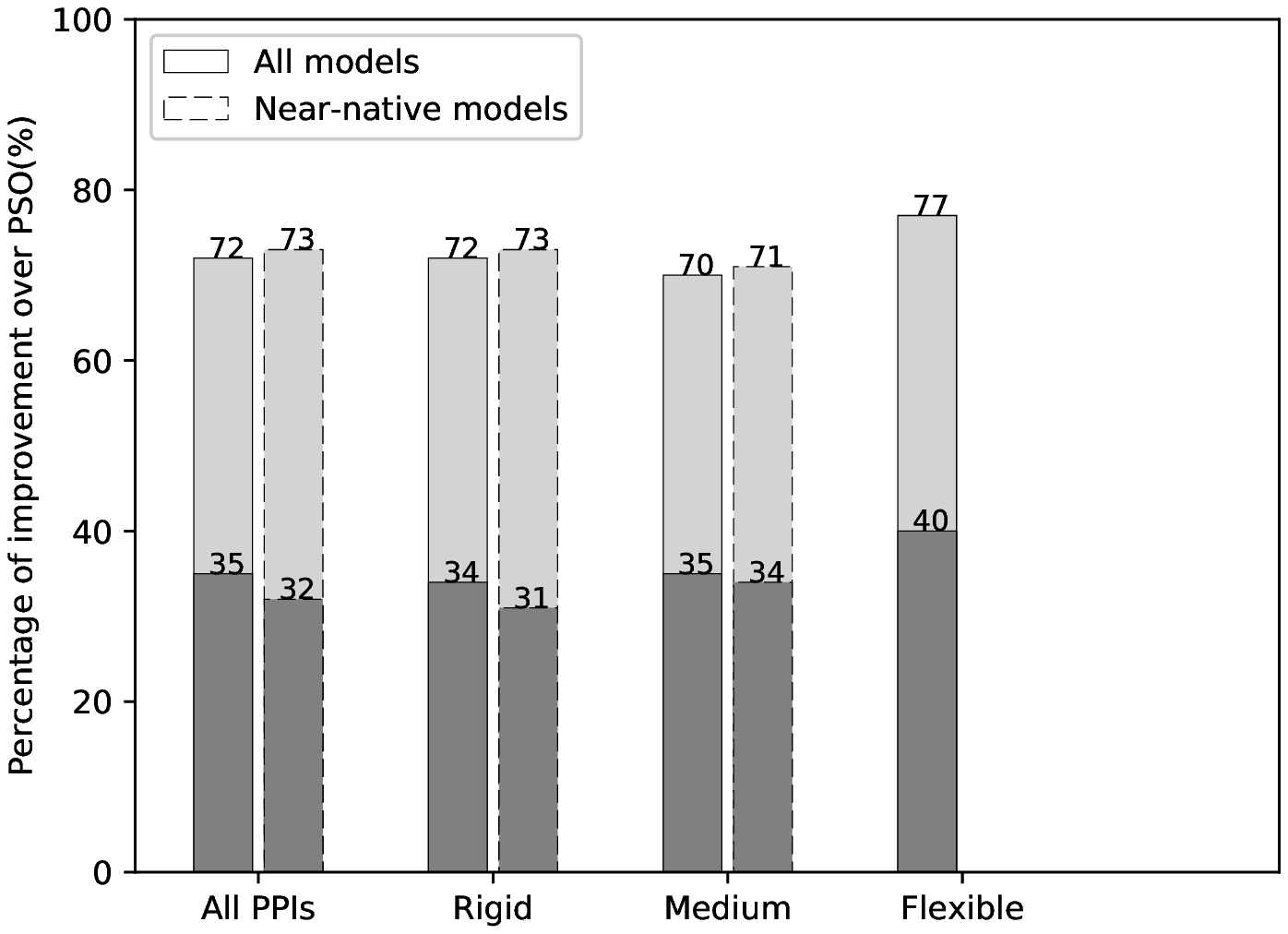}
        \caption{}
        \label{fig:3a}
    \end{subfigure}
    ~ 
    \begin{subfigure}[b]{0.25\textwidth}
        \includegraphics[width=\textwidth]{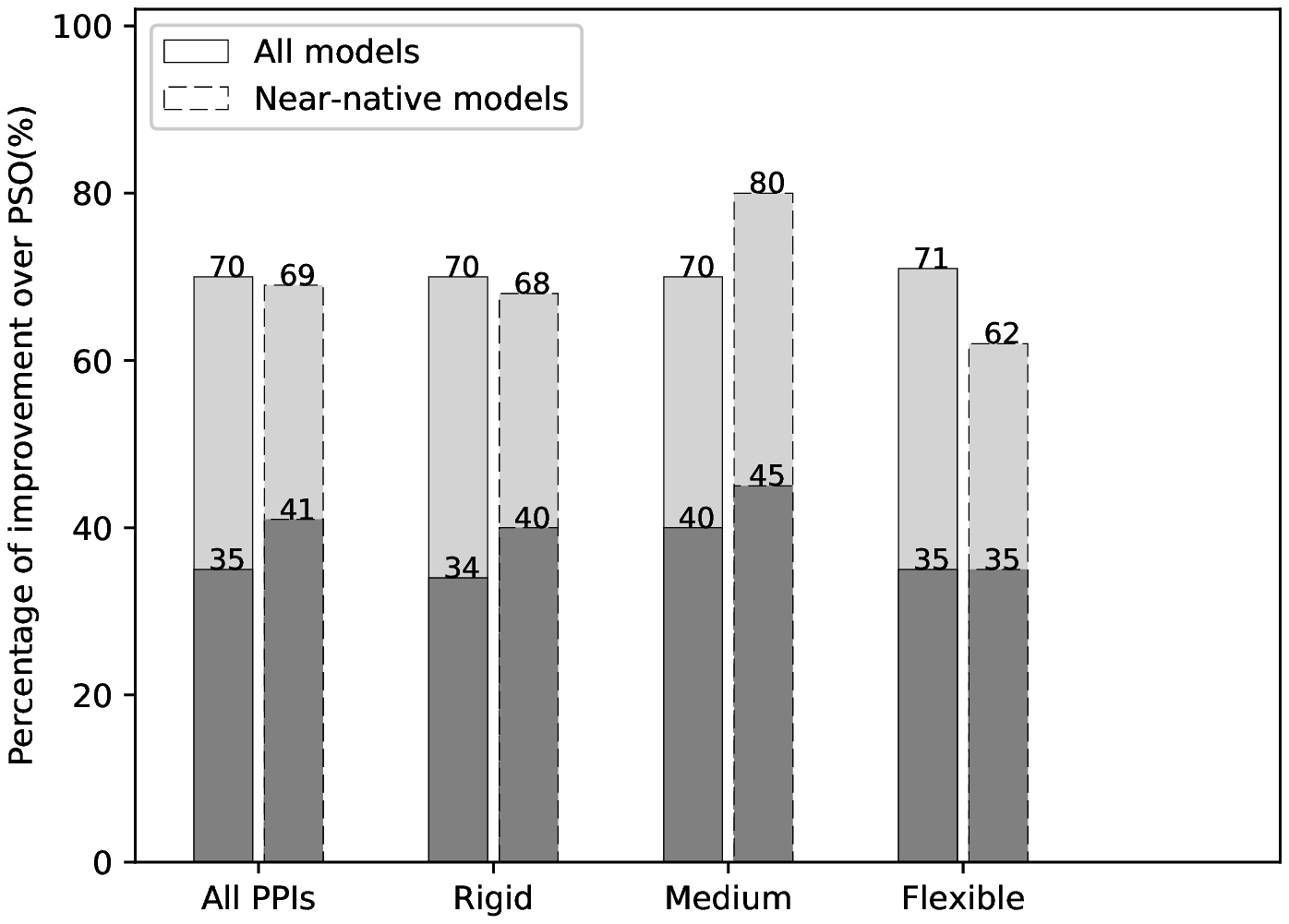}
        \caption{}
        \label{fig:3b}
    \end{subfigure}
    ~ 
    \begin{subfigure}[b]{0.25\textwidth}
        \includegraphics[width=\textwidth]{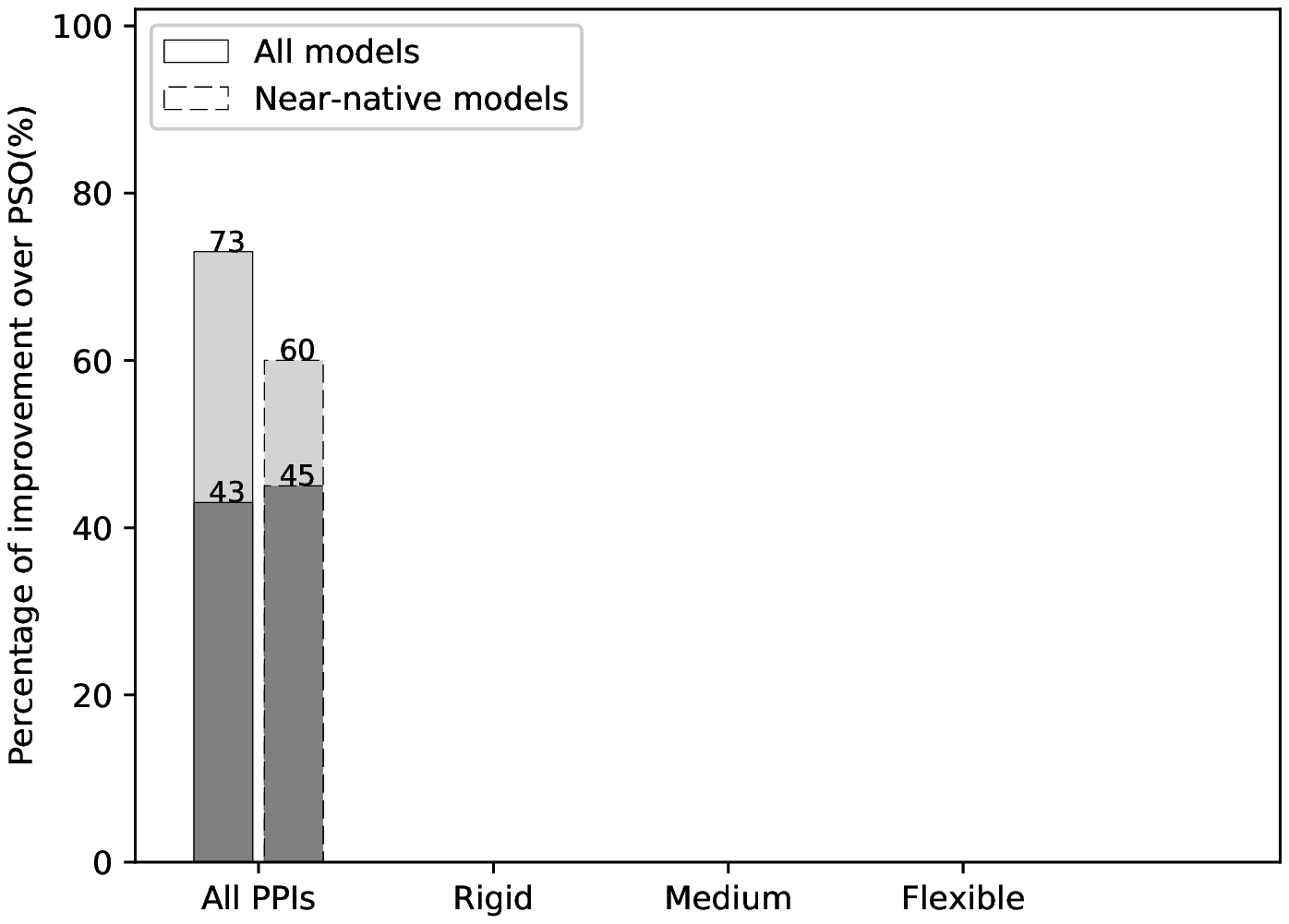}
        \caption{}
        \label{fig:3c}
    \end{subfigure}
        ~ 

    \vspace{-2em}
    \caption{The percentage of BAL predictions with iRMSD improvement against PSO for A. the training set, B. the benchmark test set, and C. the CAPRI set. The darker gray portions correspond to significant improvement (over 0.5\AA\ in iRMSD) compared to corresponding PSO predictions.}
    \label{fig:PSOvsBAL}
    \vspace{-2em}
\end{figure*}

We also compared BAL and PSO solutions head-to-head over subsets of varying difficulty levels for protein docking (Fig.\ref{fig:PSOvsBAL}).  Overall, BAL's solutions are better (or significantly better by at least 0.5\AA) than those of PSO for 70\%-80\% (31\%-45\%) of the cases, which was relatively insensitive to the docking difficulty level.  

Both PSO and BAL use a single trajectory of 31 iterations and 630 samples for each region/cluster. Most time is on local structure minimization and energy evaluations using CHARMM \citep{brooks_charmm:_2009}. The BAL running time for optimization of each cluster thus almost linearly grows with the size of the protein pair (Fig. S8), ranging from 7 hours for a 200-residue complex to 13 hours for a 1700-residue one. 

\subsection{Docking Performance: Uncertainty Quantification}

We next assess the UQ results for protein docking. Similar to that for test functions, we assess $r_{1-\sigma}$, the half length of ($1-\sigma$) confidence interval [$lb$, $ub$], i.e., $P(lb \leqslant \text{iRMSD}(\hat{\bm{x}}, \bm{x}^*) \leqslant ub) =1-\sigma$. The metrics to assess $r_{1-\sigma}$ include $\eta$, the relative error ($\eta=|\frac{r_{1-\sigma}}{|\text{iRMSD}(\hat{\bm{x}}, \bm{x}^*)}-1|$); and $\hat{P}$, the portion of the confidence intervals that actually contain the corresponding native structure across all docking runs (10 for 10 models of each protein pair in each set).

Table \ref{tab:UQ_result} shows that the portions matched well with the confidence levels over all four data sets.  Test set \textbf{b} and the CAPRI set did not have actual $K_d$ values available and were thus impacted further by the uncertainty of $K_d$ prediction, although the impact did not appear significant.  There was a trade off between the confidence level and the length of the confidence interval, as narrower confidence intervals (with less $\eta$) corresponded to lower confidence levels.  A balance seems to be at the 85\% confidence level where the relative iRMSD uncertainty is around 25\%.

%

\begin{table}
\centering
\resizebox{0.9\columnwidth}{!}{
\begin{tabular}{c|c| c c c c c }
 \hline
  Dataset& $1-\sigma$ & 0.99 & 0.95 & 0.90 & 0.85 & 0.80  \\
\hline
\multirow{2}{*}{Training}
&$\eta$ &0.40 (0.23) &
0.35 (0.18) &
0.31 (0.17) &
0.27 (0.15) &
0.22 (0.10)\\
&$\hat{P}$ & 
0.97 &
0.91 &
0.84 &
0.79 &
0.75 
  \\

\hline
\multirow{2}{*}{Test \textbf{a}}
&$\eta$ &
0.43 (0.26) &
0.39 (0.21) &
0.28 (0.16) &
0.25 (0.13) &
0.21 (0.09) 

\\
&$\hat{P}$ &
0.95 &
0.87 &
0.83 &
0.75 &
0.73
  \\
\hline
\multirow{2}{*}{Test \textbf{b}}
&$\eta$ &
0.44 (0.22)&
0.38 (0.16)&
0.26 (0.14)&
0.23 (0.10)&
0.19 (0.08) 
\\
&$\hat{P}$ & 
0.91 &
0.84 &
0.80 &
0.74 &
0.71
 \\
\hline
\multirow{2}{*}{CAPRI}
&$\eta$ &
0.43 (0.20)&
0.35 (0.13)&
0.27 (0.11)&
0.22 (0.10)&
0.20 (0.09)
\\
&$\hat{P}$ & 
0.91 &
0.85 &
0.81 &
0.74 &
0.70
  \\ \hline
\end{tabular}}
\caption{Uncertainty quantification performances of BAL on protein docking based on $\eta$, the relative error in iRMSD; and $\hat{P}$, the portion of confidence intervals from 100 runs encompassing the global optima. For $\eta$, means (and standard deviations in parentheses) are reported.}
 \label{tab:UQ_result}
 \vspace{-4em}
\end{table}

\vspace{-1em}
\subsection{Docking Performance: Quality Estimation}

For quality estimation, two metrics are used for assessing the performance. The first is Spearman's $\rho$ for ranking protein-docking predictions (structure models) for each pair.  The second is the area under the  Precision Recall Curve (AUPRC), for the binary classification of each prediction being near-native or not.  Considering that the near-natives are minorities among all predictions,  AUPRC is a more meaningful measure than the more common AUROC.  

With these two metrics we assess four scoring functions on predictions $\hat{\bm{x}}_i$: (1) $\Delta E(\hat{\bm{x}}_i)$, the MM-GBSW binding energy, i.e., the sum of the 8 features; (2) our random-forest energy model $y(\hat{\bm{x}}_i)$; (3) $P(\mathcal{X}_i|U_K)$, the conditional probability that the $i$th prediction's region is near-native give that there is at least such one in the top $K$ predictions; and (4) $P(\text{iRMSD}(\hat{\bm{x}}_i,\bm{x}^*)\leqslant 4)$, the unconditional probability that the $i$th prediction is near-native. 

\begin{figure*}[!htb]
    \centering
    \begin{subfigure}[b]{0.25\textwidth}
     \includegraphics[width=\textwidth]{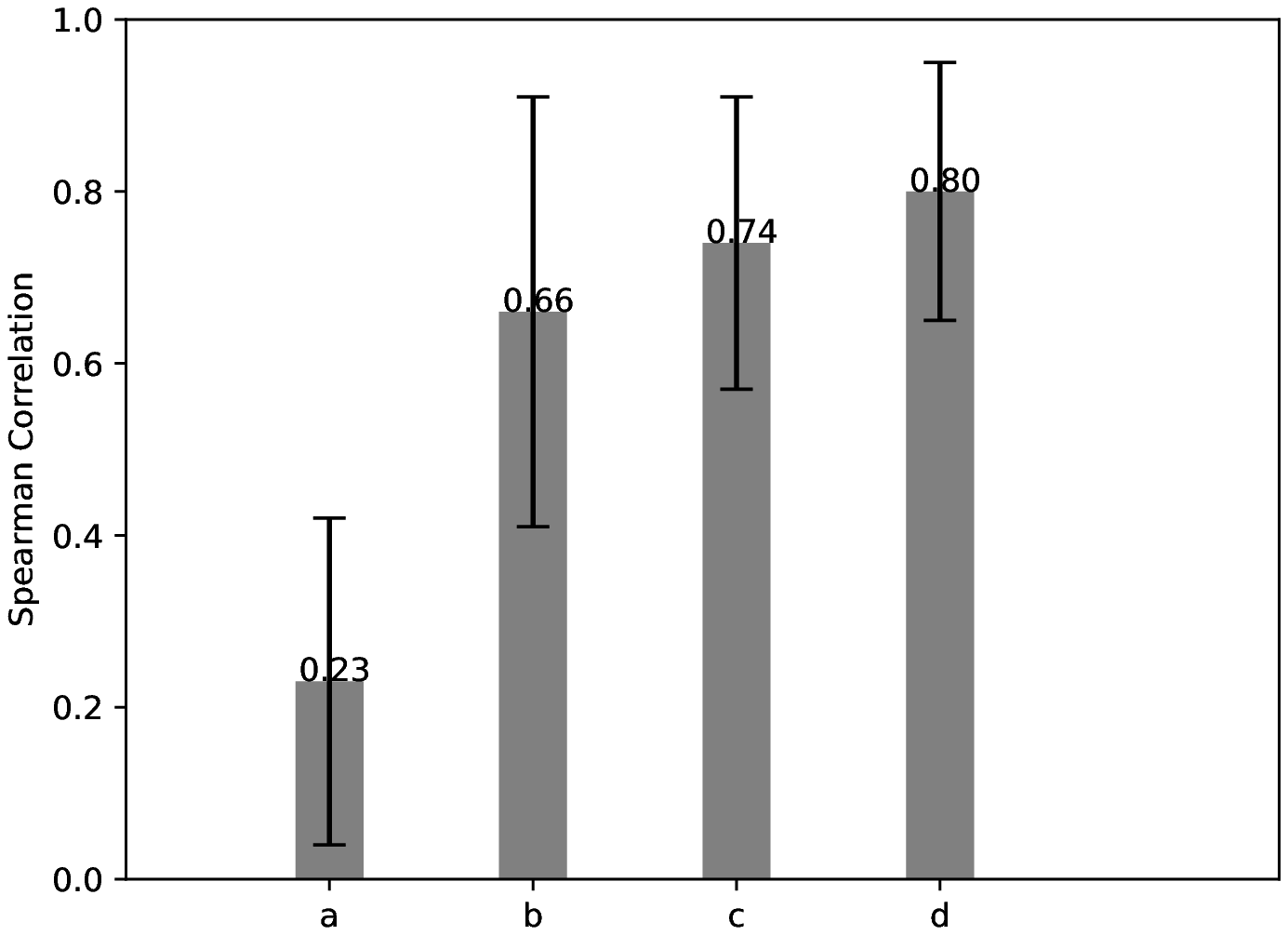}
        \caption{}
        \label{fig:gull}
    \end{subfigure}
    ~ 
    \begin{subfigure}[b]{0.25\textwidth}
       \includegraphics[width=\textwidth]{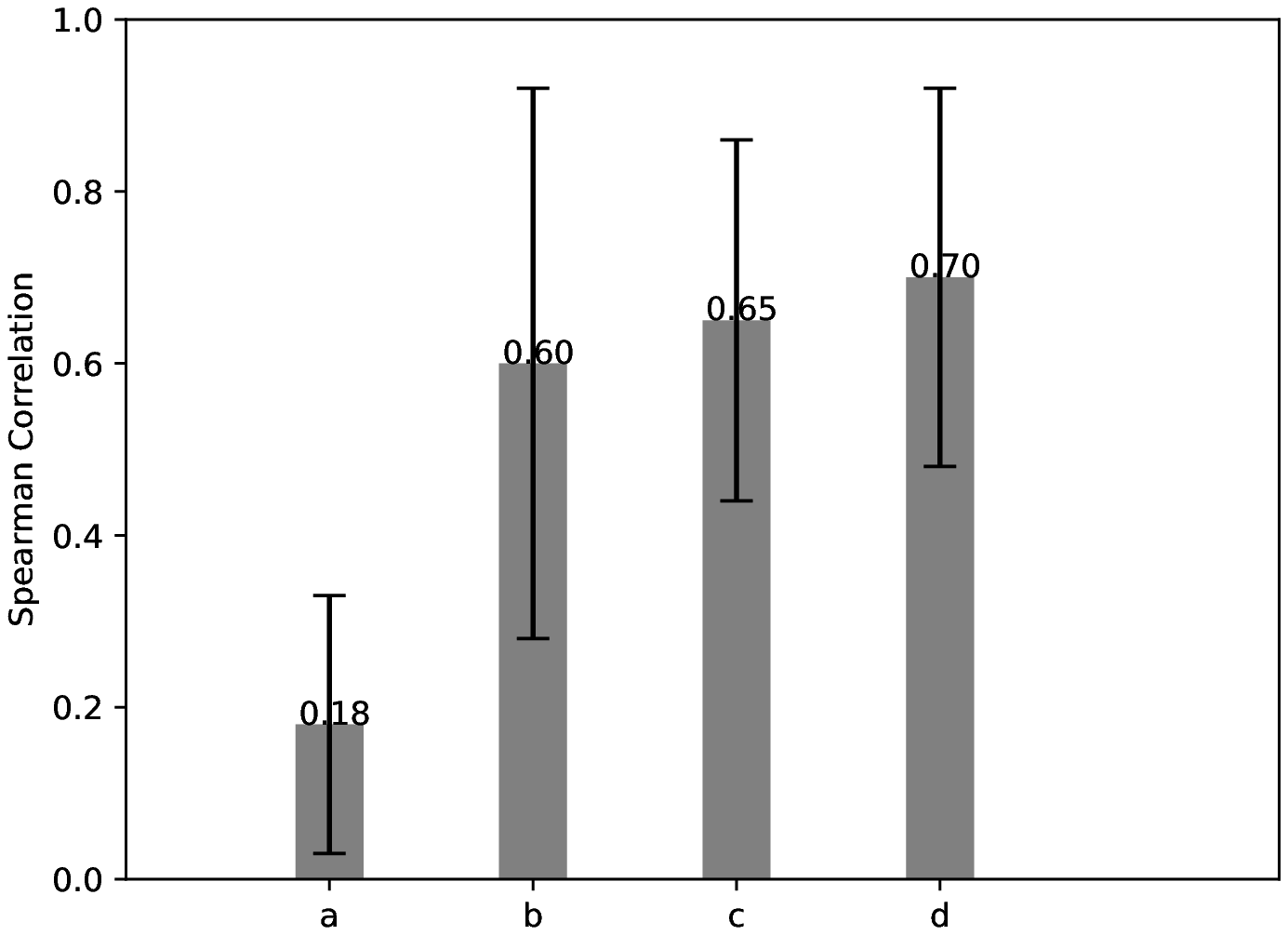}
        \caption{}
        \label{fig:tiger}
    \end{subfigure}
    ~ 
    \begin{subfigure}[b]{0.25\textwidth}
    \includegraphics[width=\textwidth]{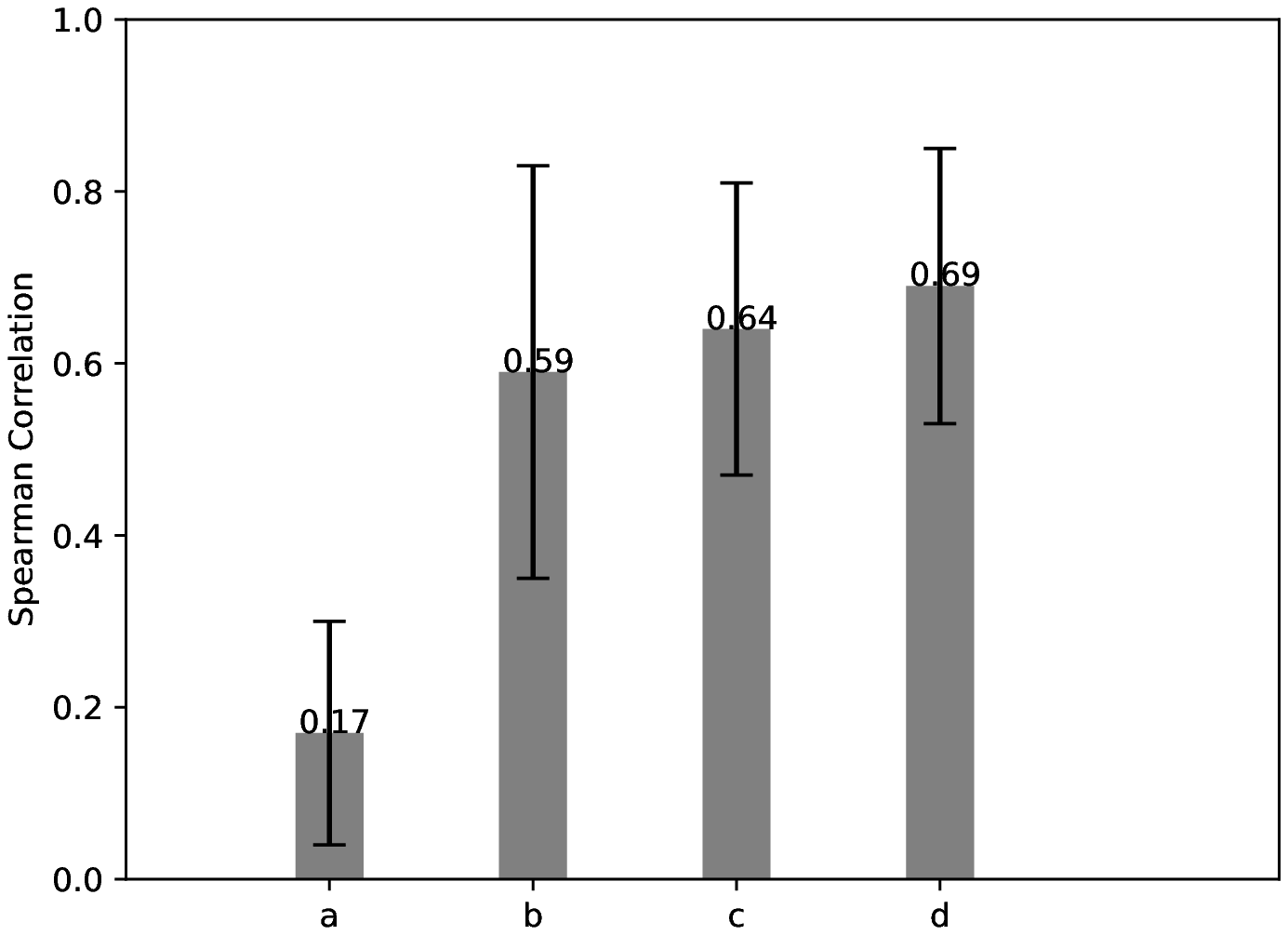}
        \caption{}
        \label{fig:mouse}
    \end{subfigure}
    \vspace{-2em}
    \caption{Ranking performance shown in the bar plot (with error bar in black) of Spearman's  $\rho$ for (A) Training set, (B) Test set and (C) CAPRI set, respectively. a,b,c,d in each figure correspond to the MM-GBSW model, our random-forest energy model, and confidence scores $P(\mathcal{X}_i|U_K)$, and $P(\text{iRMSD}(\hat{\bm{x}}_i,\bm{x}^*)\leqslant 4)$, respectively.}
    \vspace{-2em}
     \label{fig:quality_result_rank}
\end{figure*}

For ranking assessment, from Fig. \ref{fig:quality_result_rank} we find that, whereas the original MM-GBSW model merely achieved merely 0.2 for Spearman's $\rho$, our energy model using the same 8 terms as features in random forest drastically improved the ranking performance with a Spearman's $\rho$ around 0.6 for training, benchmark test, and CAPRI test sets.  Furthermore, the confidence scores $P(\mathcal{X}_i|U_K)$ and $P(\text{iRMSD}(\hat{\bm{x}}_i,\bm{x}^*)\leqslant 4)$ further improved ranking. In particular, the unconditional probability for a prediction to be near-native achieved around 0.70 in $\rho$ even for the benchmark and the CAPRI test sets.  Note that this probability, a confidence score on the prediction's near-nativeness, was derived from the posterior distribution of $\bm{x}^*$; thus it uses both enthalpic and entropic contributions.

For binary assessment on classifying the nativess of predictions, the test set is split into \textbf{a}, 26 pairs with known $K_d$ values and \textbf{b}, 100 with predicted ones.  From Table  \ref{tab:quality_result_binary} we conclude that the MM-GBSW energy model performed close to random (AUROC close to 0.5) and the random forest energy model using the same features drastically improved AUROC to around 0.8 and AUPRC $0.54\sim 0.62$ across sets.  Since AUROC is uninformative for highly imbalanced data (for instance, near-native predictions are 14\% over all data sets), we focus on AUPRC.  The next three probabilities from our BAL's confidence scores improved the AUPRC to nearly 0.80 for the training and above 0.60 for test sets.  The additional uncertainty in $K_d$ prediction from test sets  \textbf{b} and CAPRI did not noticeably impact the performance compared to test set \textbf{a}.

\begin{table}[!htb]
\centering
\resizebox{\columnwidth}{!}{
\begin{tabular}{c|c| c c c c c}
 \hline
  Dataset& Assessment & $\Delta E(\hat{\bm{x}}_i)$ & $y(\hat{\bm{x}}_i)$ & $P(\mathcal{X}_i|U_K)$ & $P(\text{iRMSD}(\hat{\bm{x}}_i,\bm{x}^*)\leqslant 4 | U_K)$ & $P(\text{iRMSD}(\hat{\bm{x}}_i,\bm{x}^*)\leqslant 4)$   \\
\hline
\multirow{2}{*}{Training}
&AUROC &
0.489&
0.806
&0.903
&0.944
&0.967
\\
&AUPRC & 
0.241&
0.624
& 0.684
& 0.771
& 0.796
\\

\hline
\multirow{2}{*}{Testing \textbf{a}}
&AUROC &
0.460 &
0.810
&0.892
&0.929
&0.939
\\
&AUPRC &
0.199&
0.550
&0.592
&0.613
&0.634

  \\
\hline
\multirow{2}{*}{Testing \textbf{b}}
&AUROC &
0.490&
0.789
& 0.847
& 0.898
& 0.927
\\
&AUPRC & 
0.203&
0.540
& 0.571
& 0.609
& 0.615

 \\
\hline
\multirow{2}{*}{CAPRI}
&AUROC &
0.491&
0.771
& 0.844
& 0.893
& 0.919

\\
&AUPRC & 
0.214&
0.561
& 0.600
& 0.610
& 0.614
  \\
  \hline
\end{tabular}
}
\caption{Binary quality estimation for our 5 scoring functions on a prediction $\hat{\bm{x}}_i$ being near-native or not: MM-GBSW energy model $\Delta E(\hat{\bm{x}}_i)$, random-forest energy model $y(\hat{\bm{x}}_i)$, and 3 BAL-determined probabilities that a region/cluster $\mathcal{X}_i$ is near-native given a native-containing list, the prediction $\hat{\bm{x}}_i$ in that region is near-native given a native-containing list, or $\hat{\bm{x}}_i$ is near-native.}
\label{tab:quality_result_binary}
\vspace{-2em}
\end{table}

\vspace{-2em}
\subsection{Docking Results by BAL Optimization and Confidence-Score Ranking}

We summarize our docking results (BAL predictions $\bm{x}_i$ ranked by confidence scores on their nativeness $P(\text{iRMSD}(\hat{\bm{x}}_i,\bm{x}^*)\leqslant 4))$ in Table \ref{tab:overall_dockingresult}; and compare them to the ZDOCK starting results (ranked by cluster size roughly reflecting entropy)  and the PSO refinement results (using the same energy model as BAL and ranked by the energy model).  We use $N_K$ to denote the number of targets with at least one near-native predictions in top $K$; and $F_K$ the fraction of such targets among all in a given set (training, benchmark test, or CAPRI test set).  
Compared to the ZDOCK starting results and PSO refinements, BAL has improved the portion of acceptable targets with top 3 predictions from  23\% and 26\%, respectively, to 32\% for the benchmark test set.   Similar improvements were found for the CAPRI set.  The portion for top 10 from BAL reached 40\% compared to ZDOCK's 33\% over the benchmark test set.  Notice that BAL only refined top 10 starting results from ZDOCK thus this improvement was purely from optimization (no ranking effect).

\begin{table*}
\centering

\setlength{\tabcolsep}{8pt}
\newcommand{\tl}{6cm}
\centering
\resizebox{0.75\textwidth}{!}{
\begin{tabular}{ r | r r r |  r r r |  r r r } 
\hline
 & \multicolumn{3}{c|}{\textbf{ZDOCK} (Starting Point)}  &  \multicolumn{3}{c|}{\textbf{PSO}} &  \multicolumn{3}{c}{\textbf{BAL}}\\
 \hline
 Dataset (size)  & $N_3$ $(F_3)$ & $N_5$ $(F_5)$ & $N_{10}$ $(F_{10})$ & $N_3$ $(F_3)$ & $N_5$ $(F_5)$ & $N_{10}$ $(F_{10})$ &  $N_3$ $(F_3)$ & $N_5$ $(F_5)$ & $N_{10}$ $(F_{10})$\\
\hline

Training (50) & 11 (22\%) & 13 (26\%) & 17 (34\%) & 15 (30\%) & 17 (34\%) & 20 (40\%) & 19 (38\%) & 20 (40\%) & 22 (44\%)\\
Test (126) & 29 (23\%) & 33 (26\%) & 41 (33\%)& 33 (26\%) & 39 (31\%) & 45 (36\%) & 40 (32\%) & 42 (33\%) & 50 (40\%)\\
CAPRI (15) & 2 (13\%) & 3 (20\%) & 4 (27\%) &  3 (20\%) & 4 (27\%) & 5 (33\%) & 4 (27\%) & 5 (33\%) & 5 (33\%)\\ \hline

\end{tabular}
}
\caption{Summary of docking results measured by the number and the portion of targets in each set that have an acceptable near-native top 3, 5, or 10 prediction.
\vspace{-2em}}
\label{tab:overall_dockingresult}
\end{table*}

\vspace{-1em}
\subsection{Energy Landscapes and Association Pathways}
\begin{figure}[hbt!]
    \centering
    \captionsetup[subfigure]{labelformat=empty}
    \begin{subfigure}[b]{0.23\textwidth}
     \includegraphics[width=\textwidth]{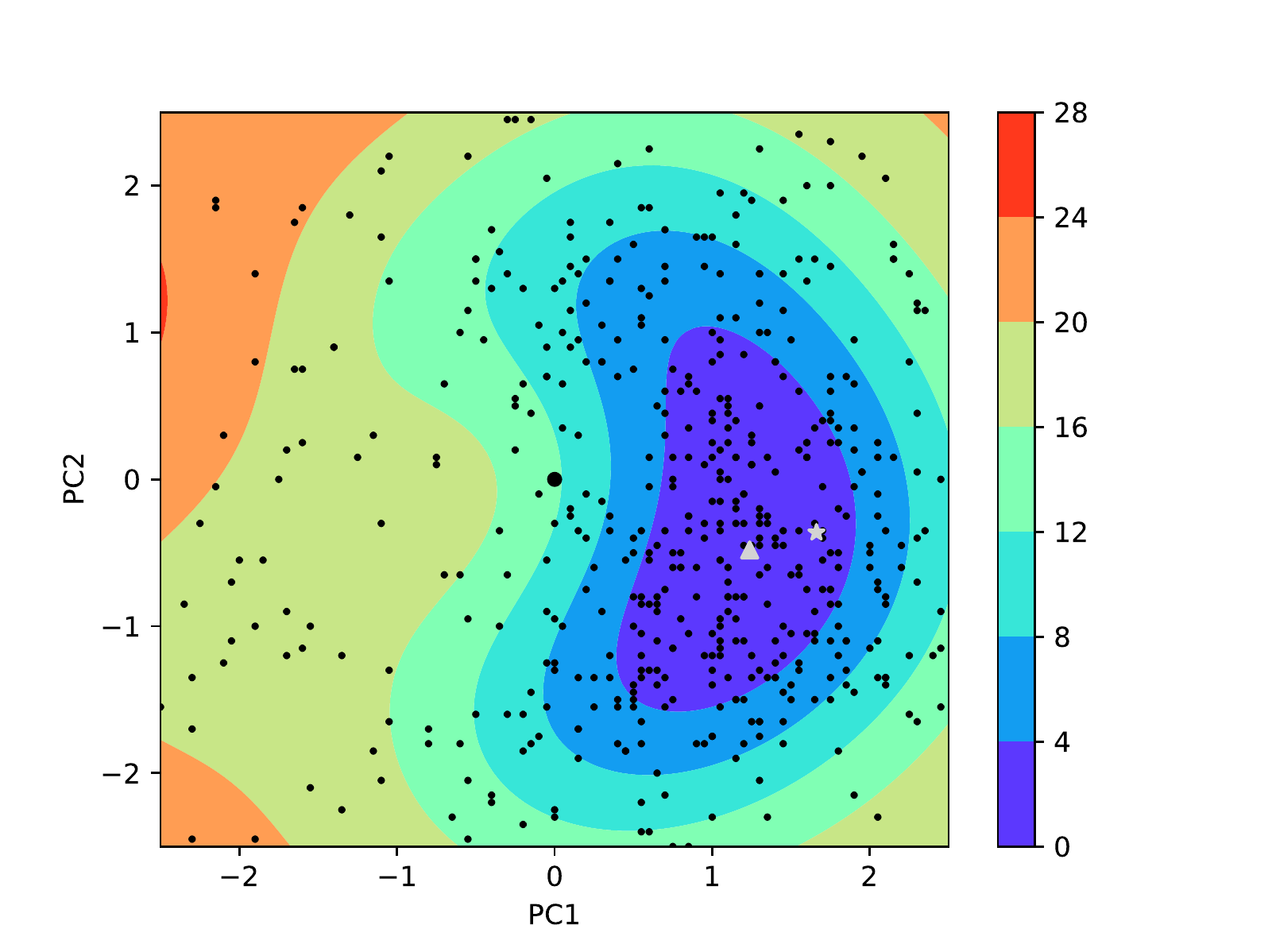}
        \caption{\textbf{4CPA\_1}}
        \label{fig:gull}
    \end{subfigure}
    ~ 
    \begin{subfigure}[b]{0.23\textwidth}
       \includegraphics[width=\textwidth]{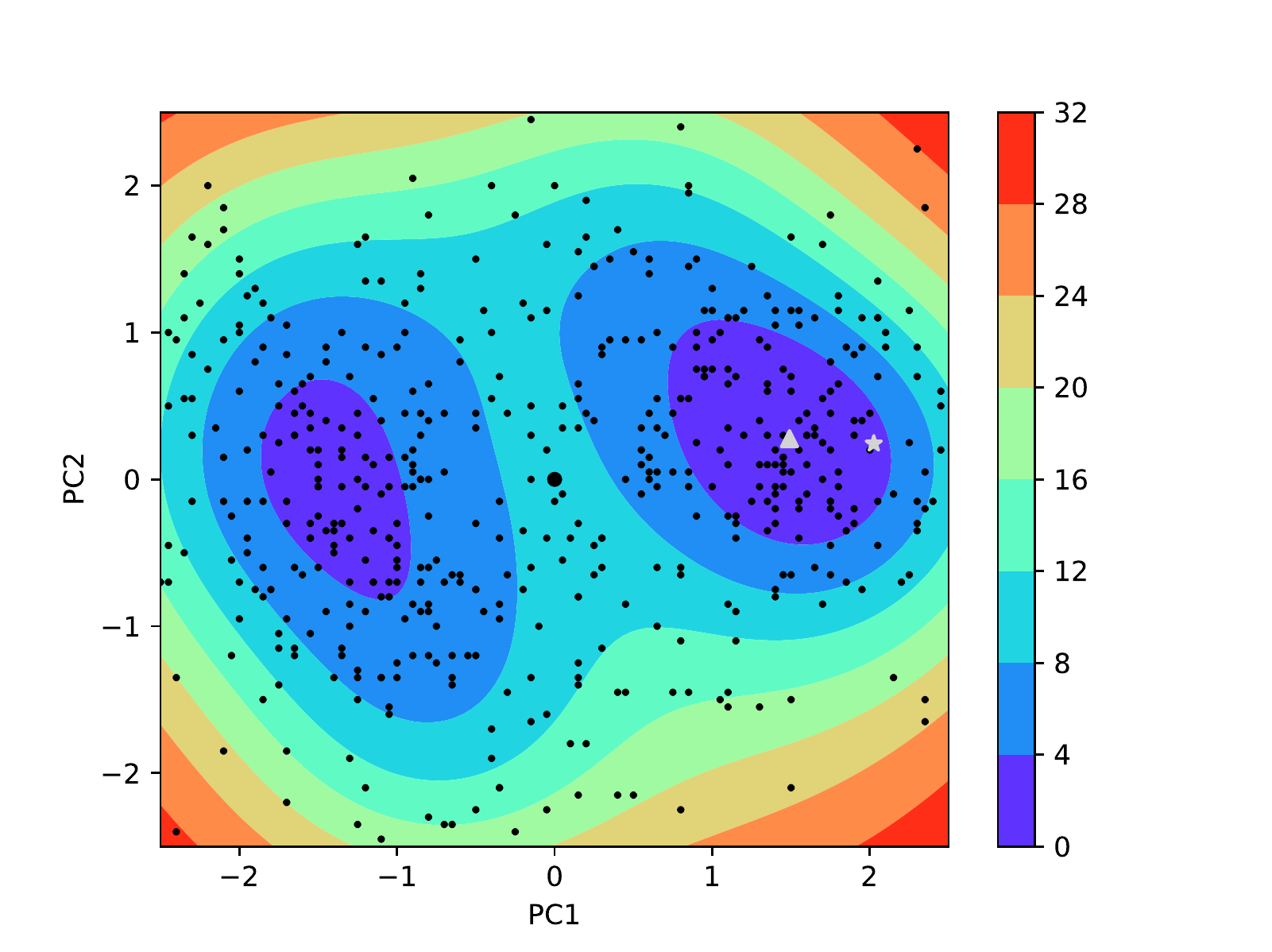}
        \caption{\textbf{1XQS\_7}}
        \label{fig:tiger}
    \end{subfigure}
    \vspace{-2em}
    \caption{The estimated energy landscapes along the first two principal components (PC) for two medium-difficulty docking cases with near-native starting models. Black dots are the samples. The grey triangle is the estimated end structure and the grey star is the true native structure. The starting structure is a thicker black dot at the origin. All the energy values are in the unit of RT and relative to the lowest sample energy value within each region.}
     \label{fig:energy_landscape}
     \vspace{-1em}
\end{figure}

We lastly investigate energy landscapes during BAL sampling. Our kriging regressor $\hat{f}(\bm{x})$ is an unbiased  estimator and works even better than the noisy observations from the random-forest energy model $y(\bm{x})$.  Energy landscapes are visualized for 37 near-native regions for the benchmark test set (Fig. S6 in the Supplementary Material) and for the CAPRI set (Fig. S7)  (non-rigid cases only).  Two examples are shown in Fig. \ref{fig:energy_landscape}, depicting a (multiple) funnel-like energy landscape with a clear association paths from the starting to the end or the native complex along gradient descents. Three more supplemental videos are provided to visualize the BAL sampling trajectories using protein structures.  
\vspace{-3em}

%

%
%
\section{Conclusions}

We present the first uncertainty quantification (UQ) study for protein docking. This is accomplished by a rigorous Bayesian framework that actively samples a noisy and expensive black-box function (i.e., collecting data $D$) while updating a posterior distribution $p(\bm{x}^* | D)$ directly over the unknown global optimum $\bm{x}^*$.  The iterative feedback between Thompson sampling and posterior updating is linked by a Boltzmann distribution with adaptive annealing schedule and non-parametric Kriging regressor.  The inverse uncertainty quantification on the location of the global optimum can easily forward-propagate for the uncertainty quantification of any quality of interest as a function of the global optimum, including the interface RMSD that measures dissimilarity between protein-docking solutions and native structures. 

We demonstrate the superb performances of Bayesian active learning (BAL) on a protein docking benchmark set as well as a CAPRI set full of homology docking.  Compared to the starting points from initial rigid docking as well as the refinement from PSO, BAL shows significant improvement, accomplishing a top-3 near-native prediction for about one-third of the benchmark and CAPRI sets. Its UQ results achieve tight uncertainty intervals whose radius is 25\% of iRMSD with a 85\% confidence level attested by empirical results.  Moreover, its estimated probability of a prediction being near-native achieves an AUROC over 0.93 and AUPRC over 0.60 (more than 4 times over random classification). 

Besides the optimization and UQ algorithms, we for the first time represent the conformational space for protein docking as a flat Euclidean space spanned by complex normal modes blending flexible- and rigid-body motions and anticipating protein conformational changes, a homogeneous and isotropic space friendly to high-dimension optimization.  We also construct a funnel-like energy model using machine learning to associate binding energies of encounter complexes sampled in docking with their iRMSD.  These innovations also contribute to the excellent performances of BAL; and lead to direct visualization of binding energy funnels and protein association pathways in conformational degrees of freedom.

\vspace{-3em}
\section*{Acknowledgements}
We thank Thom Vreven and Zhiping Weng for the ZDOCK rigid-docking decoys for the benchmark set, Haoran Chen for helping on cNMA, and  Yuanfei Sun for proofreading the manuscript.  Part of the CPU time was provided by the Texas A\&M High Performance Research Computing. 
\vspace{-2em}
\section*{Funding}
This work was supported by the National Institutes of Health (R35GM124952) and the National Science Foundation (CCF‐1546278). 

\noindent{\textit{Conflict of interest:} None declared. }

\vspace{-2em}
\small{
\bibliographystyle{natbib.bst}
\bibliography{Ref_BAL.bib}
}

\end{document}